\documentclass[pre,twocolumn,preprintnumbers,amsmath,amssymb,nofootinbib,floatfix,,superscriptaddress]{revtex4}

\usepackage{graphicx,bm}

\makeatletter
\def\graphicscale{\twocolumn@sw{0.33}{0.4}}
\def\graphicthreescale{\twocolumn@sw{0.3}{0.4}}

\begin{document}

\title{Scalar gauge-Higgs models with discrete Abelian symmetry groups}

\author{Claudio Bonati} 
\affiliation{Dipartimento di Fisica, Universit\`a di Pisa,
        and INFN, Sezione di Pisa \\ 
        Largo Pontecorvo 3, I-56127 Pisa, Italy}

\author{Andrea Pelissetto}
\affiliation{Dipartimento di Fisica, Universit\`a di Roma ``La Sapienza",
        and INFN, Sezione di Roma  \\
        P.le Aldo Moro 2, I-00185 Roma, Italy}

\author{Ettore Vicari} 
\affiliation{Dipartimento di Fisica, Universit\`a di Pisa,
        and INFN, Sezione di Pisa \\ 
        Largo Pontecorvo 3, I-56127 Pisa, Italy}

\date{\today}

\begin{abstract}
We investigate the phase diagram and the nature of the phase
transitions of three-dimensional lattice gauge-Higgs models obtained
by gauging the ${\mathbb Z}_N$ subgroup of the global ${\mathbb Z}_q$
invariance group of the ${\mathbb Z}_q$ clock model ($N$ is a
submultiple of $q$).  The phase diagram is generally characterized by
the presence of three different phases, separated by three distinct
transition lines.  We investigate the critical behavior along the two
transition lines characterized by the ordering of the scalar
field. Along the transition line separating the disordered-confined
phase from the ordered-deconfined phase, standard arguments within the
Landau-Ginzburg-Wilson framework predict that the behavior is the same
as in a generic ferromagnetic model with ${\mathbb Z}_p$ global
symmetry, $p$ being the ratio $q/N$. Thus, continuous transitions
belong to the Ising and to the O(2) universality class for $p=2$ and
$p\ge 4$, respectively, while for $p=3$ only first-order transitions
are possible. The results of Monte Carlo simulations
confirm these predictions. There is also a second transition line,
which separates two phases in which gauge fields are essentially
ordered. Along this line we observe the same critical behavior as in
the ${\mathbb Z}_q$ clock model, as it occurs in the absence of gauge
fields.
\end{abstract}

\maketitle


\section{Introduction}
\label{intro}

Classical and quantum Abelian gauge models have been extensively
studied as they provide effective theories for superconductors,
superfluids, and antiferromagnets
\cite{Fradkin_book,MM_book,RS-90,SBSVF-04,TIM-05,TIM-06,Kaul-12,KS-12,BS-13,%
  BMK-13,NCSOS-15,WNMXS-17,Sachdev-19}.  They are also supposed to
provide the effective theory for the paradigmatic example of the
quantum deconfined criticality scenario \cite{SBSVF-04}, the
transition between the N\'eel and the valence-bond-solid (VBS) state
in two-dimensional quantum antiferromagnets, see
Refs.~\cite{SBSVF-04,Kaul-12,KS-12,BS-13,BMK-13} and references
therein.  The phase diagram and the nature of the transition lines of
systems with U(1) gauge symmetry are controlled by several properties
of the model. Beside the obvious dependence on the number of
components of the scalar field, results depend on the charge $Q$ of
the scalar field
\cite{FS-79,BPPW-81,PV-19-AH3d,BPV-20-AHq2,BPV-22-AHq}, the explicit
absence or presence of monopoles \cite{PV-20-mfcp,BPV-22-mfcp}, the
compact or noncompact nature of the gauge field, see, e.g.,
Refs.~\cite{NCSOS-11,NCSOS-13,BPV-21-NCQED} and references therein.
In particular, a transition associated with a charged fixed point,
i.e., the fixed point that occurs in the corresponding field
theory~\cite{HLM-74,FH-96,IZMHS-19}, only occurs when the number of
components $N_f$ of the scalar field is sufficiently large (a numerical
study of the classical Abelian-Higgs model gives $N_f \ge 7(2)$, see
Ref.~\cite{BPV-21-NCQED}). Moreover, this type of behavior is only
observed for noncompact gauge fields or when the charge $Q$ satisfies
$Q\ge 2$ in the case of compact fields~\cite{BPV-22-AHq}.

In this work we study a different class of Abelian models, in which
the gauge group U(1) is replaced by its subgroup ${\mathbb Z}_N$.
Models with discrete local ${\mathbb Z}_2$ symmetry have been
extensively studied.  For instance, the ${\mathbb Z}_2$ gauge theory
is the paradigmatic example of a model undergoing a topological
transition, without a local order
parameter~\cite{Wegner-71,Kogut-79,Sachdev-19}, and is often used as a
toy model to understand nonperturbative properties of lattice gauge
models relevant for high-energy physics, see, e.g.,
Refs.~\cite{GR-02,GGRT-03}. Moreover, they are relevant to interpret
critical transitions in magnetic
systems~\cite{SF-00,MSF-01,SSS-02,PD-05,Nussinov-05}, in liquid
crystals~\cite{LRT-93,LNNSWZ-15}, and in models relevant for quantum
computations (this is the case of the ${\mathbb Z}_2$ gauge theory
coupled with an Ising
system~\cite{VDS-09,TKPS-10,WDP-12,SSN-21,Grady-21,HSAFG-21,BPV-21-Z2Higgs}).

To define the model that we consider, we start from the ${\mathbb
  Z}_q$ clock model, in which the scalar fields are phases that take
$q$ discrete values, and we gauge the ${\mathbb Z}_N$ subgroup of the
invariance symmetry group ${\mathbb Z}_q$.  Models with global
${\mathbb Z}_q$ symmetry occur in several contexts and have attracted
significant interest in recent years because of their connection with
the N\'eel-VBS transition in antiferromagnets, see
Refs.~\cite{SSNHS-03,HS-03,PSS-20,SGS-20} and references therein.  In
the absence of gauge fields, for $q\ge 4$, one observes the phenomenon
of symmetry enlargement at the transition. Large-scale universal
properties become O(2) invariant, the ${\mathbb Z}_q$ anisotropy
playing the role of a dangerously irrelevant operator. It is important
to stress that models with discrete gauge and global symmetry groups
are also relevant in view of their possible realization using
cold-atom quantum technology.  Indeed, in this framework it is
essential that the Hilbert space be finite.  Possible implementations
of ${\mathbb Z}_2$ gauge systems have recently been proposed, see,
e.g., Ref.~\cite{BSADGG-19}.  Moreover, the discretization of the
scalar degrees of freedom leads to faster classical and quantum
computations~\cite{Hasenbusch-19,Hasenbusch-20,KS-19,JLZ-20}.

In this work we wish to investigate the role that the gauge symmetry
group plays in determining the phase diagram and the nature of the
transition lines. We find that all transition lines where the matter
fields show long-range correlations can be interpreted in terms of an
effective Landau-Ginzburg-Wilson theory in which one only considers
the dynamics of a scalar order parameter. As a consequence, the
universality class of the transitions only depends on the global
symmetry group and on the discrete nature of the scalar field. A
second, important issue is the question of the symmetry enlargement at
the transition. In other words, we would like to determine under which
conditions one can observe the O(2) critical behavior in the presence
of ${\mathbb Z}_N$ local invariance.

The paper is organized as follows.  In Sec.~\ref{sec2} we define the
model, while in Sec.~\ref{sec3} we specify the quantities that are
determined in the Monte Carlo simulations.  The expected phase diagram
is discussed in Sec.~\ref{sec4}, while the numerical results are
presented in Sec.~\ref{sec5}.  Finally, in Sec.~\ref{sec6} we
summarize the results and draw our conclusions.  In the Appendices we
report some useful results. In App.~\ref{App.A} and \ref{App.B} we
report exact results for the ${\mathbb Z}_4$ and ${\mathbb Z}_8$
models with ${\mathbb Z}_2$ gauge invariance. In App.~\ref{App.C} we
compute the relevant scaling functions for the Ising and the XY model
that are compared with the numerical results in Sec.~\ref{sec5}.

\section{The model} \label{sec2}

We consider a ${\mathbb Z}_N$ gauge model coupled with a complex
scalar field defined on a cubic lattice.  The fundamental fields are
complex phases $w_{\bm x}$, satisfying $|w_{\bm x}| = 1$, associated
with the sites of the lattice and phases $\sigma_{{\bm x},\mu}$,
$|\sigma_{{\bm x},\mu}|=1$, associated with the lattice links. These
phases can only take $q$ and $N$ values, respectively, where $q$ is an
integer multiple of $N$. More precisely, we set
\begin{equation}
w = \exp (2 \pi i m /q)\,, \qquad \sigma = \exp (2 \pi i n /N)\,,
\label{defwsigma}
\end{equation}
where $m = 0,\ldots, q-1$, $n = 0,1\ldots N-1$. 

The corresponding Hamiltonian is defined as 
\begin{equation}
    H = H_{\rm kin} + H_{\rm g}\,.
\label{Hamiltonian}
\end{equation}
The first term is 
\begin{equation}
H_{\rm kin} = - J\ \hbox{Re} \sum_{{\bm x}, \mu} \bar{w}_{\bm x}
\sigma_{{\bm x},\mu}\, {w}_{{\bm x}+\hat\mu}\,,
\label{hcpnla}
\end{equation}
where the sum is over all lattice sites ${\bm x}$ and directions $\mu$
($\hat{\mu}$ are the corresponding unit vectors). The second term is 
\begin{equation}
H_{\rm g} = - g \sum_{{\bm x},\mu>\nu} \hbox{Re }\Pi_{{\bm x},\mu\nu}\,,
\end{equation}
where the sum is over all lattice plaquettes, and the plaquette
contribution is given by
\begin{equation}
\Pi_{{\bm x},\mu\nu} = 
   \sigma_{{\bm x},\mu} \sigma_{{\bm x} + \hat{\mu},\nu}
   \bar{\sigma}_{{\bm x} + \hat{\nu},\mu} \bar{\sigma}_{{\bm x},\nu}\,, 
\end{equation}
The partition function is 
\begin{equation}
Z = \sum_{z,\sigma} e^{-H/T}\,.
\label{Z-model}
\end{equation}
In the following we use $\beta = J/T$ and $\kappa = g/T$ as
independent variables.  The model is invariant under local ${\mathbb
  Z}_N$ and global ${\mathbb Z}_{q} $ transformations. The global
symmetry group is ${\mathbb Z}_{q}/{\mathbb Z}_N = {\mathbb Z}_p$ with
\begin{equation}
    p = {q\over N}\,.
\end{equation}
The model is well defined also if $N$ is unrelated to $q$, but in this
case it is only invariant under ${\mathbb Z}_M$ local transformations,
where $M$ is the greatest common divisor of $N$ and $q$. If $N$ is an
integer multiple of $q$, the model is invariant under local ${\mathbb
  Z}_q$ transformations and it is possible to integrate out the scalar
fields, performing the change of variable ${\tau}_{{\bm x},\mu} =
{\sigma}_{{\bm x},\mu} \bar{w}_{{\bm x}} w_{{\bm x} + \hat{\mu}}$.
One obtains the Hamiltonian of a ${\mathbb Z}_N$ gauge model in the
presence of a linear gauge-symmetry breaking term,
\begin{equation}
H = - g \sum_{{\bm x},\mu>\nu} \Pi_{{\bm x},\mu\nu} - J\ \hbox{Re}
\sum_{{\bm x}, \mu} \tau_{{\bm x},\mu}\,.
\end{equation}
where the plaquette is expressed in terms of the new field $\tau$.
For $N=q=2$ this model has been extensively
studied~\cite{GR-02,GGRT-03,Nussinov-05,VDS-09,TKPS-10,WDP-12,SSN-21,%
  Grady-21,HSAFG-21,BPV-21-Z2Higgs}. Here we shall focus on the case
$q > N$.

\section{The observables} \label{sec3}

In our numerical study we consider cubic lattices of linear size $L$.
As we are dealing with topological transitions, one should carefully
choose the boundary conditions. We consider open boundary conditions,
to avoid slowly-decaying dynamic modes that are present in systems
with periodic boundary conditions. Indeed, in the latter case, the
Polyakov loops (the product of the gauge compact fields along
nontrivial lattice paths that wrap around the lattice) have a very
slow dynamics, especially in the gauge deconfined phase, if one uses
algorithms with local updates. For open boundary conditions, Polyakov
loops are not gauge invariant and thus their dynamics is not relevant
for the estimation of gauge-invariant observables. A local algorithm
is therefore efficient.  Of course, open boundary conditions give rise
to additional scaling corrections, due to the boundary, and thus
larger systems are needed to obtain asymptotic results.

We simulate the system using a standard Metropolis algorithm.  We
compute the energy densities and the specific heats
\begin{eqnarray}
&E_k = {1\over V} \langle H_{\rm kin}\rangle\,,\quad &C_k ={1\over
    V} \left( \langle H^2_{\rm kin} \rangle - \langle H_{\rm kin}
  \rangle^2\right)\,, \nonumber \\ & E_g = {1\over V} \langle H_{\rm
    g} \rangle\,,\quad &C_g ={1\over V} \left( \langle H^2_{\rm g}
  \rangle - \langle H_{\rm g} \rangle^2\right)\,,
\label{ecvdef}
\end{eqnarray}
where $V=L^3$.

We consider the two-point correlation function of the field $w$ with
charge $Q$:
\begin{eqnarray}
G_Q({\bm x},{\bm y}) &= &
   \hbox{Re} \, 
   \langle (\bar{w}_{\bm x} {w}_{\bm y})^Q \rangle\,,
\label{gxyp} \\
&=& \left\langle \cos \left(2 \pi Q (m_{\bm x} - m_{\bm y})/q
\right)\right\rangle\,.  \nonumber
\end{eqnarray}
where $m_{\bm x}$ is defined in Eq.~(\ref{defwsigma}).  If $Q$ is a
multiple of $N$, the correlation function $G_Q({\bm x},{\bm y})$ is
gauge invariant.  Then, we define the Fourier transform
\begin{equation}
\widetilde{G}_Q({\bm p}) = 
    {1\over V} \sum_{{\bm x},{\bm y}} 
      e^{i{\bm p}\cdot ({\bm x} - {\bm y})} G_Q({\bm x},{\bm y})
\end{equation}
($V$ is the volume), and the corresponding susceptibility and 
correlation length,
\begin{eqnarray}
&&\chi_Q =  
\widetilde{G}_Q({\bm 0})\,, 
\label{chisusc}\\
&&\xi^2_Q \equiv  {1\over 4 \sin^2 (\pi/L)}
{\widetilde{G}_Q({\bm 0}) - \widetilde{G}_Q({\bm p}_m)\over 
\widetilde{G}_Q({\bm p}_m)}\,,
\label{xidefpb}
\end{eqnarray}
where ${\bm p}_m =
(2\pi/L,0,0)$. Note that, since we use open boundary conditions, the choice 
of ${\bm p}_m$ is somewhat arbitrary. Other choices, as long as they satisfy 
$|p_m| \sim 1/L$, would be equally valid.

In our FSS analysis we use renormalization-group invariant 
quantities. We consider
\begin{equation}
R_{\xi,Q} = \xi_Q/L\,,
\end{equation}
and the charge-$Q$ Binder parameter
\begin{equation}
U_Q = {\langle \mu_{2,Q}^2\rangle \over \langle \mu_{2,Q}\rangle^2}\,, \qquad
\mu_{2,Q} = \sum_{{\bm x}{\bm y}}
   \hbox{Re} \, 
   (\bar{w}_{\bm x} {w}_{\bm y})^Q\,,
\label{binderdef}
\end{equation}
To determine the nature of the transition, one can consider the $L$
dependence of the maximum $C_{\rm max}(L)$ of one of the specific
heats.  At a first-order transition, $C_{\rm max}(L)$ is proportional
to the volume $L^3$, while at a continuous transition it behaves as
\begin{equation}
C_{\rm max}(L) = a L^{\alpha/\nu} + C_{\rm reg} \,.
\end{equation}
The constant term $C_{\rm reg}$, due to the analytic background, is
the dominant contribution if $\alpha < 0$.  The analysis of the
$L$-dependence of $C_{\rm max}(L)$ may allow one to distinguish
first-order and continuous transitions.  However, experience with
models that undergo weak first-order transitions indicates that in
many cases the analysis of the specific heat is not conclusive. The
maximum $C_{\rm max}(L)$ may start scaling as $L^3$ at values of $L$
that are much larger than those at which simulations can be actually
performed.  A more useful quantity is a Binder parameter $U$, which
has a qualitatively different behavior at continuous and first-order
transitions. In the latter case, the maximum $U_{\rm max}(L)$ of $U$
at fixed size $L$ increases with the volume~\cite{CLB-86,VRSB-93}. On
the other hand, $U$ is bounded as $L\to \infty$ at a continuous phase
transition.  In this case, in the FSS limit, any renormalization-group
invariant quantity $R$ scales as
\begin{eqnarray}
&&R(\beta,L) \approx  f_R(X) + L^{-\omega} f_{c,R}(X)\,, \label{rsca}\\
&&X = (\beta-\beta_c) L^{1/\nu} \,,\nonumber
\end{eqnarray}
where $\omega$ is a correction-to-scaling exponent.  Thus, a
first-order transition can be identified by verifying that $U_{\rm
  max}(L)$ increases with $L$, without the need of explicitly
observing the linear behavior in the volume.

In the case of weak first-order transitions, the nature of the
transition can also be understood from the combined analysis of $U$
and $R_{\xi}$. At a continuous transition, in the FSS limit any
renormalization-group invariant quantity $R$ scales as
\begin{equation}
R(\beta,L) = F_R(R_\xi) + L^{-\omega} F_{c,R}(R_\xi) + \ldots\,,
\label{r12sca}
\end{equation}
where $F_R(x)$ is universal and $F_{c,R}(x)$ is universal apart from a
multiplicative constant. The Binder parameter $U$ does not obey this
scaling relation at first-order transitions, because of the divergence
of $U$ for $L\to\infty$.  Therefore, the order of the transition can
be understood from plots of $U$ versus $R_\xi$.  The absence of a data
collapse is an early indication of the first-order nature of the
transition.

\section{Predicted phase diagram} \label{sec4}

\begin{figure}
\includegraphics*[width=6cm,angle=0]{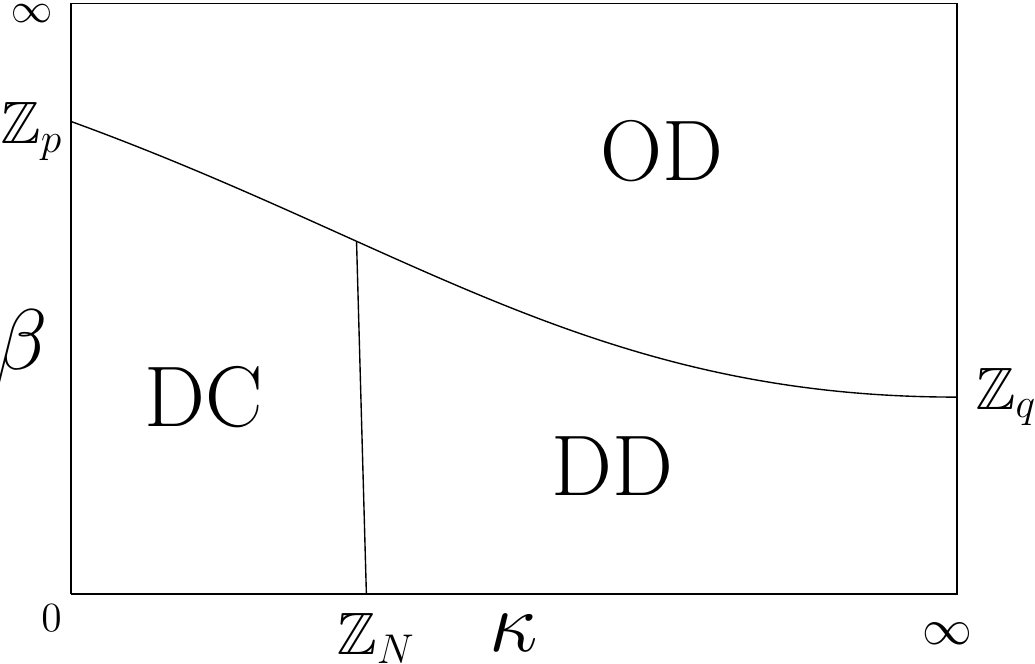}
\caption{Phase diagram of the model. Three phases are present: a
  disordered-confined (DC) phase, a disordered-deconfined (DD) phase,
  and an ordered-deconfined (OD) phase. For $\kappa = 0$ there is a
  ferromagnetic ${\mathbb Z}_{p}$ ($p=q/N$) transition, for $\beta =
  0$ a topological ${\mathbb Z}_N$ transition, and for $\kappa \to
  \infty$ a ferromagnetic ${\mathbb Z}_{q}$ transition.  }
\label{phdia}
\end{figure}

Our simulations are consistent, as we shall see in Sec.~\ref{sec5},
with the phase diagram shown in Fig.~\ref{phdia}, with three different
phases.  To clarify their nature and the universality class of the
different transition lines, it is useful to discuss some limiting
cases.

In the limit $\kappa\to\infty$, the gauge degrees of freedom freeze
and one can set $\sigma_{{\bm x},\mu} = 1$ on all links (when open
boundary conditions are used this is also true in a finite volume),
obtaining the ferromagnetic ${\mathbb Z}_q$ clock model, which
undergoes a standard finite-$\beta$ transition. For $q=2$ and 4 it belongs
to the Ising universality class, for $q=3$ it is of first order, while
for $q\ge 5$ the critical behavior is the same as in the XY model, see
Refs.~\cite{HS-03,Hasenbusch-19,PSS-20}. 

Note that a ${\mathbb Z}_q$
perturbation is irrelevant~\cite{CPV-03} at the XY fixed point for any
$q\ge 4$ and, in particular, also for $q=4$.  Thus, XY critical
behavior is generically expected in models with ${\mathbb Z}_4$ global
invariance and it has been indeed observed in systems with soft
${\mathbb Z}_4$ breaking~\cite{PAD-15,SGS-20}.
The standard ${\mathbb Z}_4$ clock model, which undergoes an Ising
transition~\cite{HS-03}, is an exception. It behaves differently,
because the model can be formulated in terms of two decoupled Ising
spins on each site. In generic ${\mathbb Z}_4$ systems with discrete
fields, one can still parametrize the model in terms of two Ising
spins, but now they are coupled by an energy-energy interaction.  At
the decoupled Ising fixed point, this perturbation is relevant,
although with a rather small renormalization-group
dimension~\cite{CPV-03} given by $2/\nu_{\rm Is} - 3 = 0.17475(2)$, if
we use the estimate $\nu_{\rm Is} = 0.629971(4)$ of the Ising-model
exponent \cite{KPSV-16}. The energy-energy interaction is the one that
drives the system towards the XY fixed point, if the transition is
continuous.

For $\beta = 0$, there are no scalar fields and one obtains a pure
gauge ${\mathbb Z}_N$ model, that can be related by
duality~\cite{SSNHS-03} to a ${\mathbb Z}_N$ spin model, with a global
${\mathbb Z}_N$ symmetry. The ${\mathbb Z}_N$ gauge theory undergoes a
topological transition at $\kappa_c$, which belongs to the same
universality class as the corresponding transition in the ${\mathbb
  Z}_N$ spin clock model.  Estimates of $\kappa_c$ can be found in
Ref.~\cite{BCCGPS-14}.  For $N=2$, we can use the results of
Ref.~\cite{FXL-18} for the standard Ising model and duality to
estimate $\kappa_c = 0.761413292(12)$. For $N\to \infty$, one
has~\cite{BCCGPS-14,BPV-22-AHq}
\begin{equation}
\kappa_c\simeq \kappa_{gc} \, N^2,
\end{equation}
where $\kappa_{gc}=0.076051(2)$ is the critical coupling of the
inverted XY model~\cite{NRR-03}.

For $\kappa = 0$, one can sum over the gauge fields and obtain a
gauge-invariant Hamiltonian that depends on the fields $w_{\bm x}$
only. For generic values of $N$, the expression of the effective
Hamiltonian is complex and not very illuminating. However, one expects
the same critical behavior for any local Hamiltonian that is invariant
under local ${\mathbb Z}_N$ transformations.  One such Hamiltonian is
\begin{eqnarray}
H_{\rm eff} &=& - J \sum_{{\bm x},\mu} \hbox{Re } 
     (\bar{w}_{{\bm x} + \hat\mu} w_{\bm x})^N  \nonumber \\
     &=& - J \sum_{{\bm x},\mu} 
   \cos\left[ {2 \pi N\over q} 
        (m_{{\bm x} + \hat{\mu}} - m_{{\bm x}}) \right]\,.
\label{HZp}
\end{eqnarray}
If we express $m_{\bm x} = p n_{1,\bm x} + n_{2,\bm x}$, with
$n_{1,\bm x} = 0,\ldots N-1$, $n_{2,\bm x} = 0,\ldots p-1$ ($p=q/N$),
we obtain the Hamiltonian of a ferromagnetic ${\mathbb Z}_{p}$ clock
model (the Ising model for $p=2$). Analogously, the correlation
function $G_N({\bm x},{\bm y})$ in the model with Hamiltonian
(\ref{HZp}) is equivalent the correlation function $G_1({\bm x},{\bm
  y})$ in the ${\mathbb Z}_{p}$ clock model.  For $N=2$ and $q=4$, one
can show that the Hamiltonian (\ref{Hamiltonian}) is exactly
equivalent to Eq.~(\ref{HZp}) for $\kappa = 0$, see App.~\ref{App.A},
and thus, in this case, the relation of the gauge model with the Ising
model is exact.

On the basis of the previous argument we predict the universality
class of the transition at $\kappa=0$ to depend only on the ratio $p =
q/N$, and to be the same as that of the ${\mathbb Z}_p$ clock model (as
we discuss below, for $p\not=4$).  Therefore, if the transitions are
continuous, they should belong to the Ising universality class for
$p=2$ and to the XY universality class for $p\ge 5$. For $p=3$,
instead, we expect a discontinuous transition as in the ${\mathbb
  Z}_3$ clock model. For $p=4$, the transition in the ${\mathbb Z}_4$
clock model is not the generic one expected in ${\mathbb Z}_4$
invariant systems.  In App.~\ref{App.B} we have performed an exact
calculation for the ${\mathbb Z}_8$ model with ${\mathbb Z}_2$ gauge
invariance.  For $\kappa = 0$, the model can be rewritten in terms of
two Ising spins $\rho_{\bm x}^{(1)}$ and $\rho_{\bm x}^{(2)}$, with
Hamiltonian
\begin{eqnarray}
H_{\rm eff} &=& \sum_{{\bm x},\mu} [
    B(\beta)  \rho_{\bm x}^{(1)} \rho_{{\bm x}+\hat{\mu}}^{(1)} + 
    B(\beta)  \rho_{\bm x}^{(2)} \rho_{{\bm x}+\hat{\mu}}^{(2)}  \nonumber \\
&& \quad 
    + C(\beta)  \rho_{\bm x}^{(1)} \rho_{{\bm x}+\hat{\mu}}^{(1)}
              \rho_{\bm x}^{(2)} \rho_{{\bm x}+\hat{\mu}}^{(2)}]\,,
\end{eqnarray}
where the functions $B(\beta)$ and $C(\beta)$ can be derived using the
results of App.~\ref{App.B}. The decoupling of the two Ising systems
that occurs in the ${\mathbb Z}_4$ clock model does not occur here
[$C(\beta)$ does not vanish] and therefore we expect XY behavior, if
the transition is continuous. The same result is expected for any $q$
and any $N$, with $p=4$.

Finally, several results~\cite{LNNSWZ-15} are available for $N=2$ in
the limit $q\to \infty$, in which $w_{\bm x}$ is an unconstrained
phase and the global invariance group is U(1).  The phase diagram is
similar to the one reported in Fig.\ref{phdia}. There are three
different phases that can be characterized by the behavior of the
gauge and scalar degrees of freedom~\cite{LNNSWZ-15}.  For small
$\beta$ and $\kappa$, gauge modes are confined, while they are
deconfined in the other two phases. As for the scalar degrees of
freedom, they are disordered in the two small-$\beta$ phases, while
they are ordered (the $\mathbb{Z}_q$ symmetry is broken) in the
large-$\beta$ phase. The three phases are separated by three
transition lines.  Along the transition lines that separate the
large-$\beta$ ordered-deconfined (OD) phase from the two low-$\beta$
phases, transitions belong~\cite{LNNSWZ-15} to the XY universality
class, as in the models obtained for $\kappa \to 0$ and $\kappa \to
\infty$. Along the line that separates the disordered-confined (DC)
phase from the disordered-deconfined (DD) phase one expects the same
behavior as in the ${\mathbb Z}_N$ gauge model obtained for $\beta =
0$.

It is conceivable, and we shall verify it in the next section, that
the same phase diagram holds for the models we consider here as long
as $q > N$.  Moreover, as in the case $q=\infty$, we expect the
critical behavior along the three lines to be the same as at the
corresponding endpoint at $\beta = 0$, $\kappa = 0$, and
$\kappa\to\infty$. The only exceptions might occur for $N=4$, along
the DC-DD line and for $q=4$ along the DD-OD line. In these cases, it is
{\em a priori} possible to observe XY behavior instead of Ising
behavior. However, since the crossover exponent of the relevant
perturbation that drives the system out of the decoupled Ising fixed
point is rather small~\cite{CPV-03}, significant crossover effects may
be present.

\section{Numerical results} \label{sec5}

\subsection{Small-$\kappa$ transition line }

Let us now discuss the behavior along the DC-OD transition line. For
this purpose we have performed simulations at fixed $\kappa$, varying
$\beta$.  In all cases, we set $\kappa = 0.4$, which, on the basis of
the estimates of $\kappa_c$ at $\beta = 0$ reported in
Ref.~\cite{BCCGPS-14}, should guarantee that we are studying a
transition belonging to the DC-OD line.  As we mentioned, we expect
the phase behavior to depend only on $p = q/N$.

\subsubsection{Models with $p=2$}

\begin{figure}[tbp]
\includegraphics[scale=\graphicscale,angle=-90]{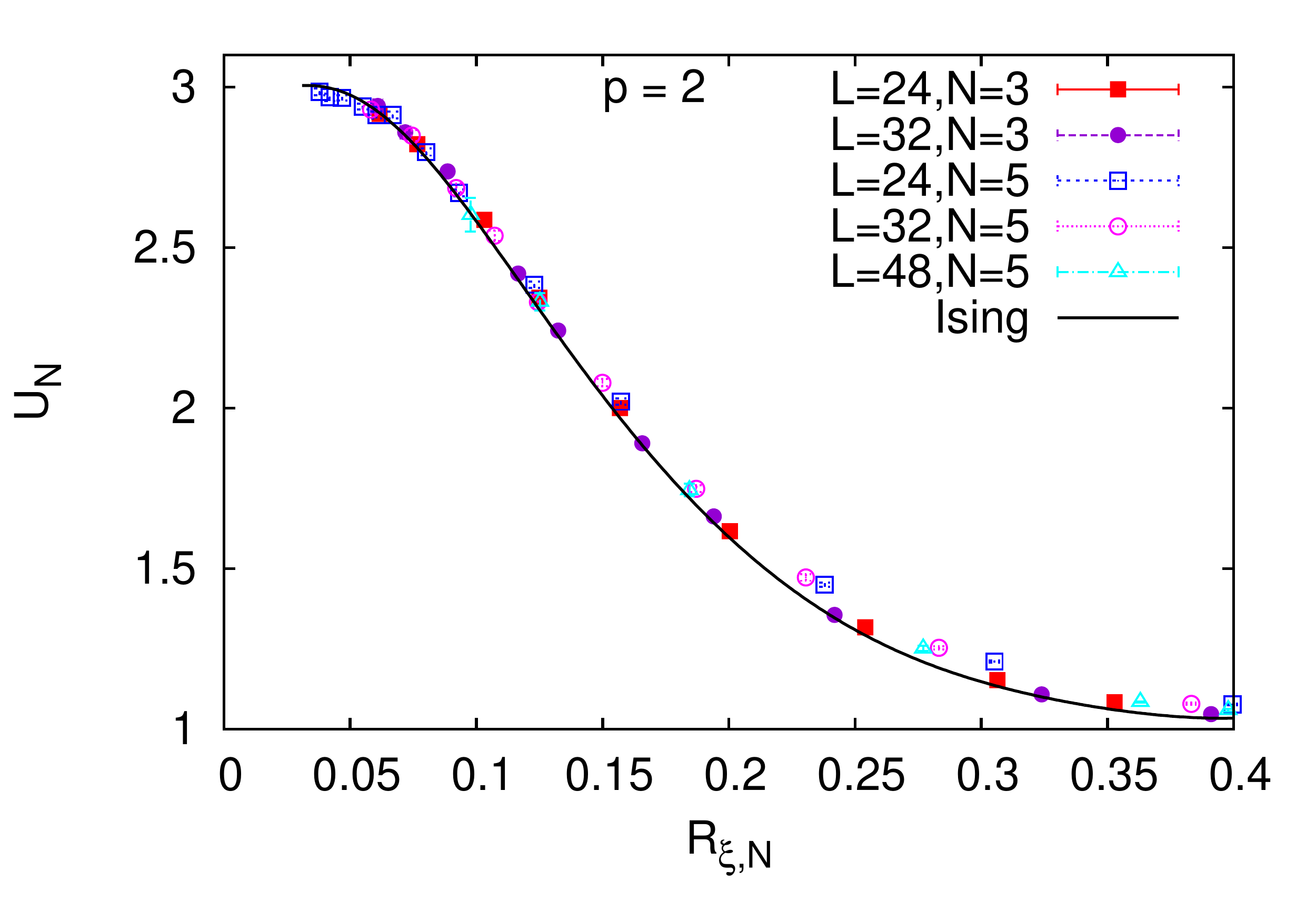}
\caption{Estimates of $U_N$ versus $R_{\xi,N}$ for the models with
  $N=3$, $q=6$ and with $N=5$, $q = 10$, at fixed $\kappa = 0.4$. In
  both cases $p = q/N = 2$. We also report the universal curve
  $F_{\xi,1}(R_{\xi,N})$ computed in the Ising model (``Ising").}
\label{URxi-p2-kappa0p4}
\end{figure}

\begin{figure}[tbp]
\includegraphics*[scale=\graphicscale,angle=-90]{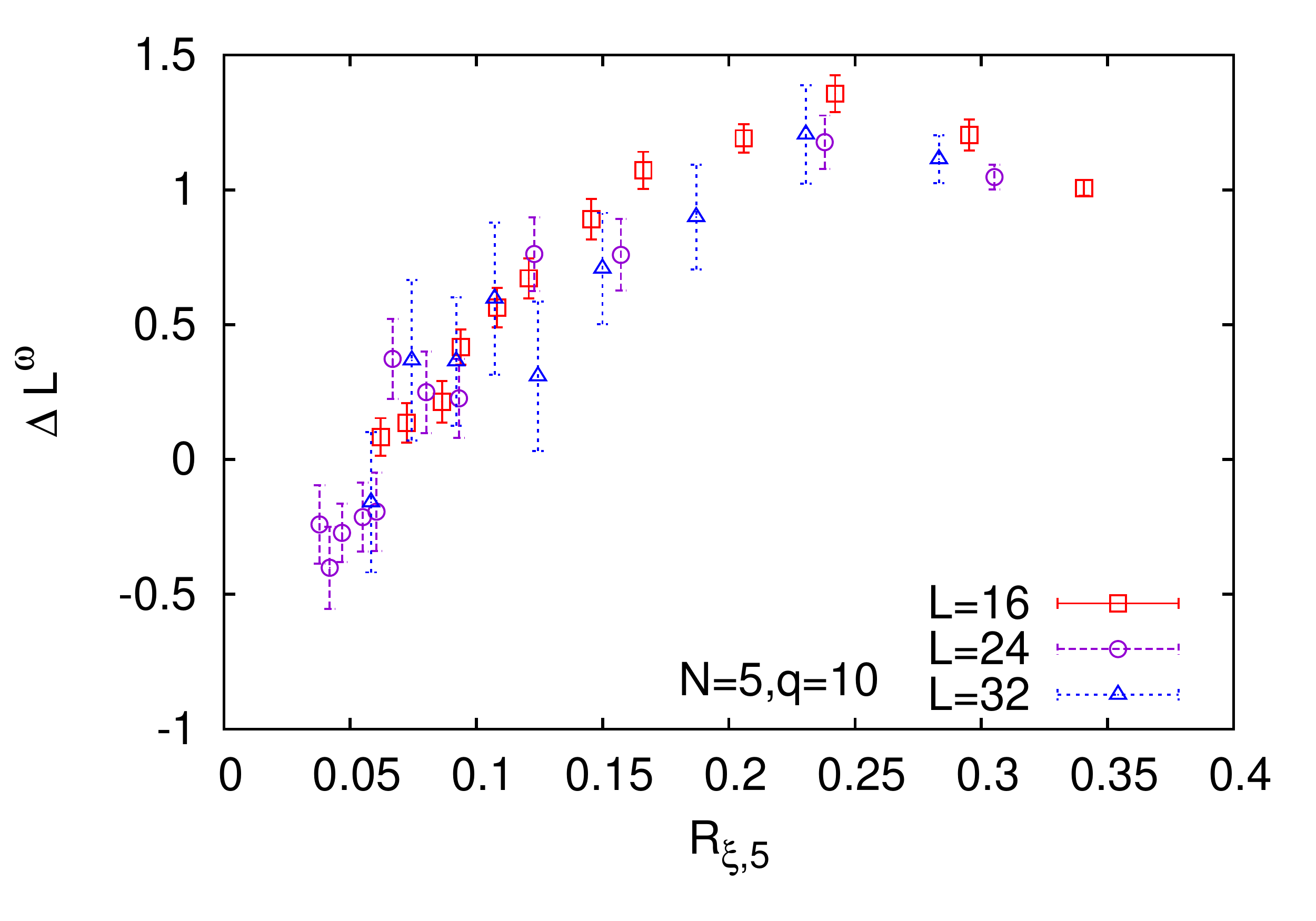}
\caption{Estimates of $L^{\omega}\Delta$ versus $R_{\xi,N}$ for $N=5$,
  $q = 10$ ($p = q/N = 2$), at fixed $\kappa = 0.4$. The function
  $\Delta$ is defined in Eq.~(\ref{deviations-def}). We use the
  correction-to-scaling exponent for Ising systems, $\omega = 0.83$.
}
\label{DevURxi-p2-kappa0p4}
\end{figure}

For $p=2$ we have performed simulations for $(q,N) = (4,2), (6,3)$ and
$(10,5)$. In Fig.~\ref{URxi-p2-kappa0p4} we report we report $U_N$
versus $R_{\xi,N} = \xi_N/L$ for $N=3$ and 5, together with the
scaling function $F_{U,1}(R_{\xi,N})$, where $F_{U,1}(x)$ is the
asymptotic scaling function that expresses $U_1$ in terms of
$R_{\xi,1}$ in the Ising model (the computation is discussed in
App.~\ref{App.C}).  For $N=3$, the data essentially fall on top of the
Ising scaling curve, while the results for $N=5$ show tiny deviations
that decrease as $L$ increases. To provide a better check that the
asymptotic behavior for $N=5$ is the same as in the Ising model, we
have determined the corrections, defining
\begin{equation}
    \Delta(R_{\xi,N}) = U_N - F_{U,1}(R_{\xi,N})\,.
 \label{deviations-def}
\end{equation}
If the transition belongs to the Ising universality class, the
estimates of $L^\omega \Delta(R_{\xi,N})$ should approximately belong
to a single curve, provided one uses $\omega = 0.8303(18)$, which is
the correction-to-scaling exponent for the Ising universality
class~\cite{EPPSSV-12-14,KPSV-16}. The results are shown in
Fig.~\ref{DevURxi-p2-kappa0p4}. For $N=5$, $q=10$ we observe a nice
scaling.  Moreover, as expected, the shape of the curve is similar to
that observed for the Ising model, see App.~\ref{App.C}.

As a additional check, we have performed combined fits of $U_N$ and
$R_{\xi,N}$ to Eq.~(\ref{rsca}), parametrizing $f_R(x)$ and
$f_{c,R}(x)$ with polynomials. If we let $\nu$ be a free parameter and
fix $\omega = 0.83$ (the value for the Ising universality class), we
obtain $\nu = 0.62(1)$ ($N=3$) and $0.64(1)$ ($N=5$).  These results
are consistent with the Ising prediction $\nu =
0.629971(4)$~\cite{EPPSSV-12-14,KPSV-16}.  To estimate the position of
the critical point, we have then performed fits fixing $\nu =
0.629971$. We obtain $\beta_c = 1.4546(1)$ and $\beta_c = 4.5660(7)$
for $N=3$ and $N=5$, respectively.

\begin{figure}[tbp]
\includegraphics*[scale=\graphicscale,angle=-90]{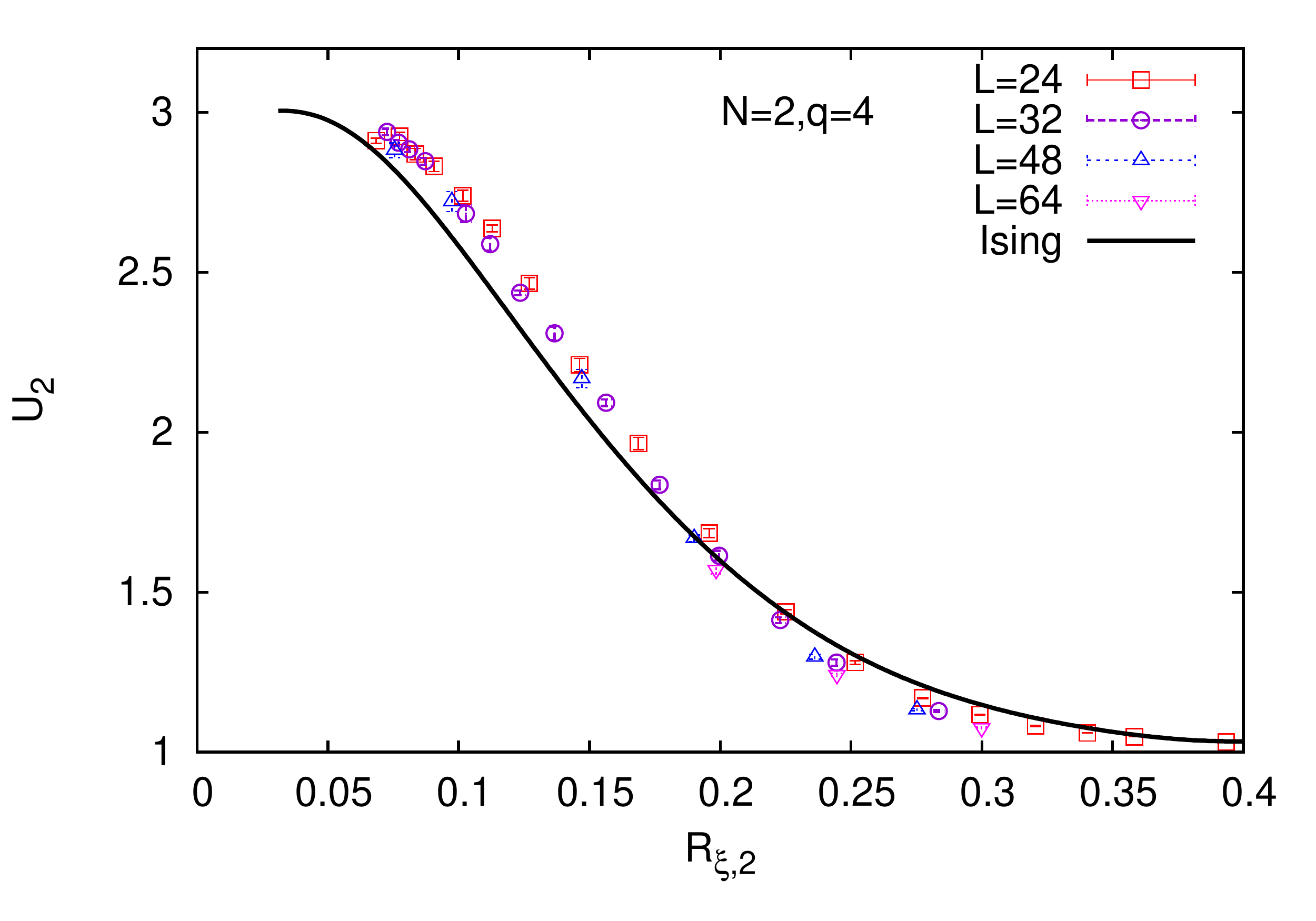}
\caption{Estimates of $U_2$ versus $R_{\xi,2}$ for the model with
  $N=2$, $q=4$ at fixed $\kappa = 0.4$. Results are compared with the
  scaling function computed in the Ising model, as in
  Fig.~\ref{URxi-p2-kappa0p4}.}
\label{URxi-N2q4-kappa0p4}
\end{figure}

\begin{figure}[tbp]
\includegraphics*[scale=\graphicscale,angle=-90]{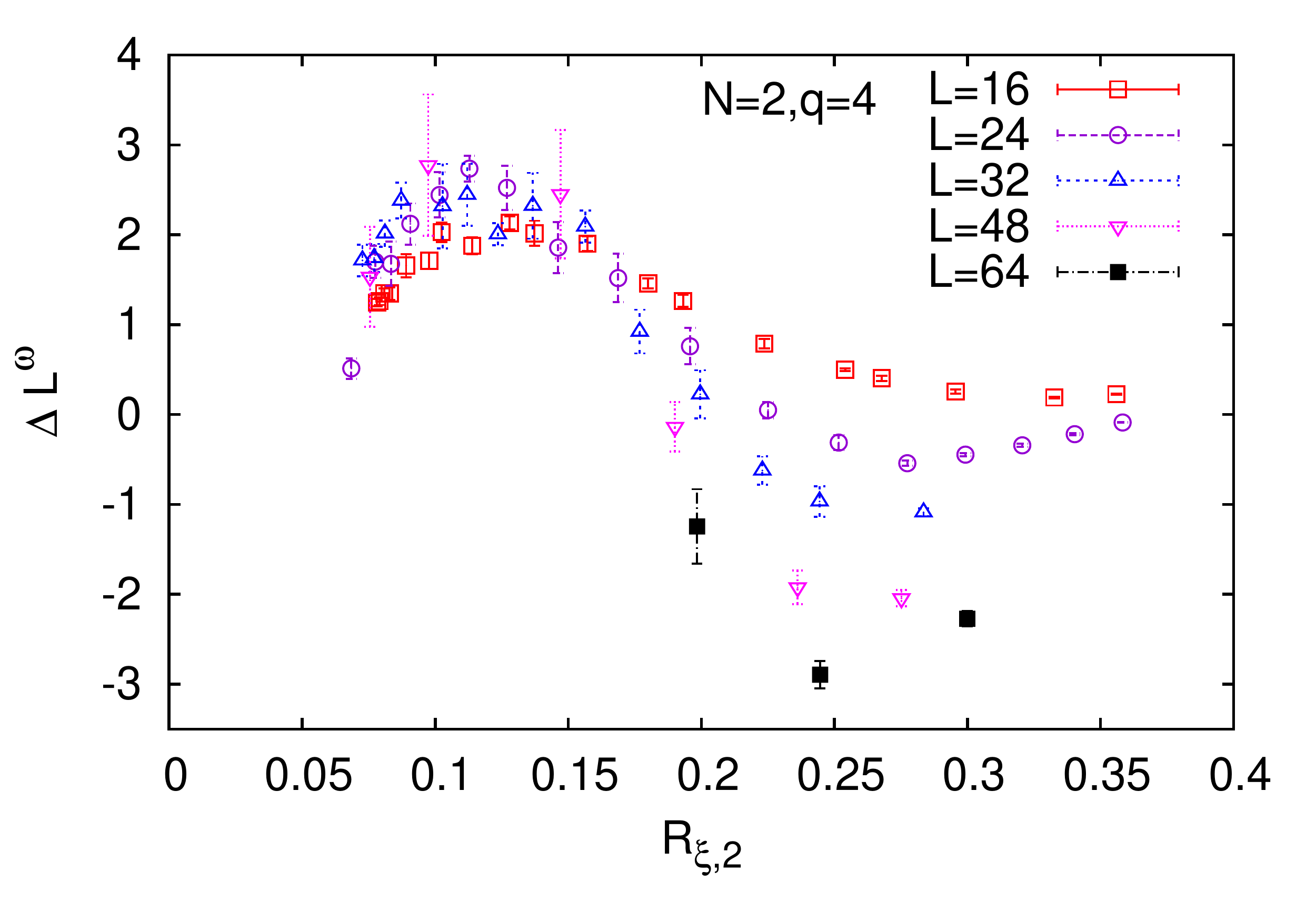}
\caption{Estimates of $L^{\omega}\Delta$ versus $R_{\xi,2}$ for $N=2$,
  $q=4$ at fixed $\kappa = 0.4$.  The function $\Delta$ is defined in
  Eq.~(\ref{deviations-def}). We use the correction-to-scaling
  exponent for Ising systems, $\omega = 0.83$.  }
\label{DevURxi-N2q4-kappa0p4}
\end{figure}

Finally, we consider the case $N=2$ and $q=4$. The estimates of $U_2$
versus $R_{\xi,2}$ are reported in Fig.~\ref{URxi-N2q4-kappa0p4}.
Data are close to the Ising curve. However, at a closer look,
deviations from the Ising curve do not decrease as $L$ increases. This
is evident from Fig.~\ref{DevURxi-N2q4-kappa0p4}, where we report the
deviations from the Ising curve. For $0.2\lesssim R_{\xi,2} \lesssim
0.35$ deviations apparently increase as $L$ increases.

\begin{figure}[tbp]
\includegraphics*[scale=\graphicscale,angle=-90]{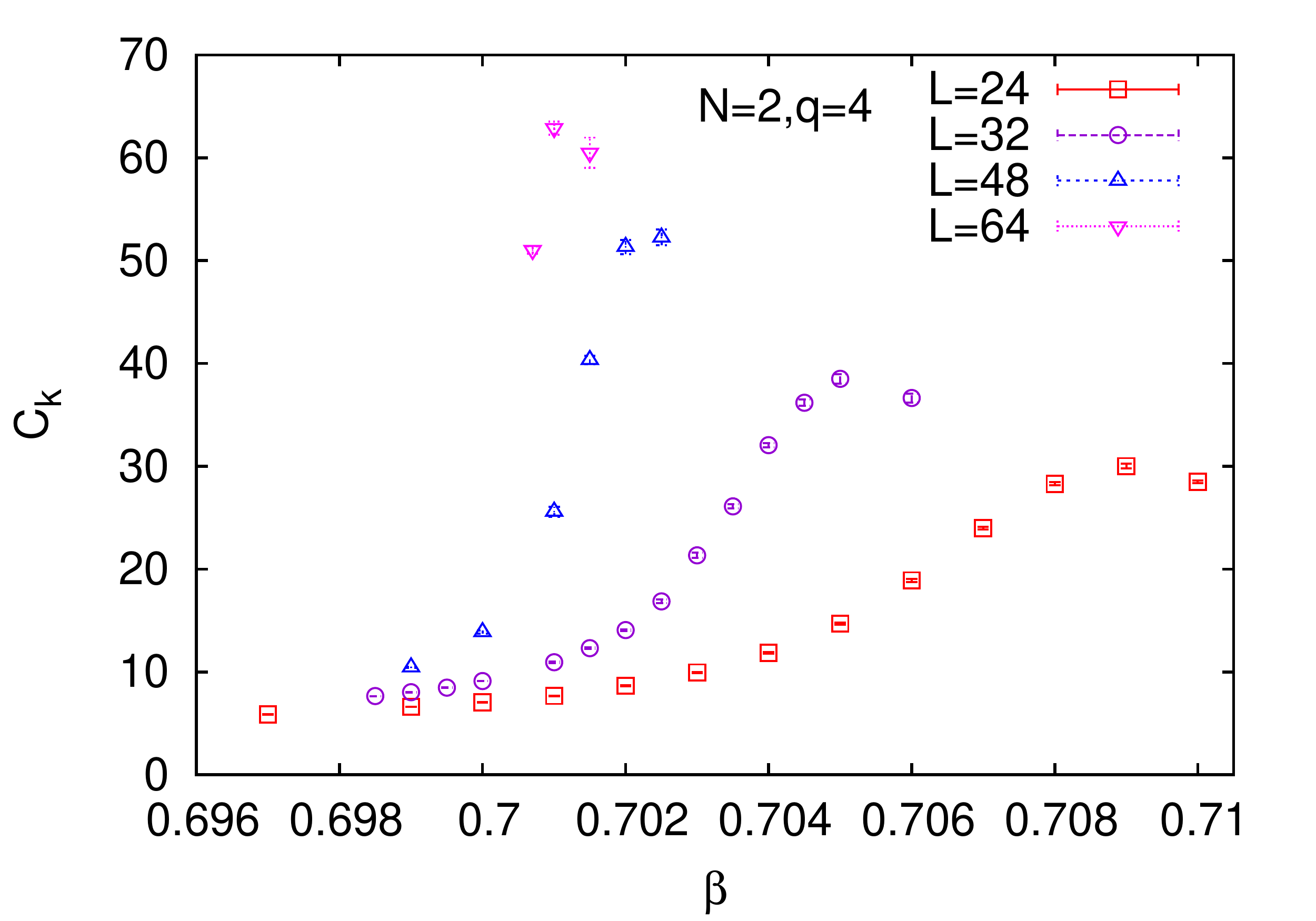}
\caption{Estimates of the specific heat $C_k$ versus $\beta$ for
  $N=2$, $q=4$ at fixed $\kappa = 0.4$.  }
\label{Ck-N2q4-kappa0p4}
\end{figure}

\begin{figure}[tbp]
\includegraphics*[scale=\graphicscale,angle=-90]{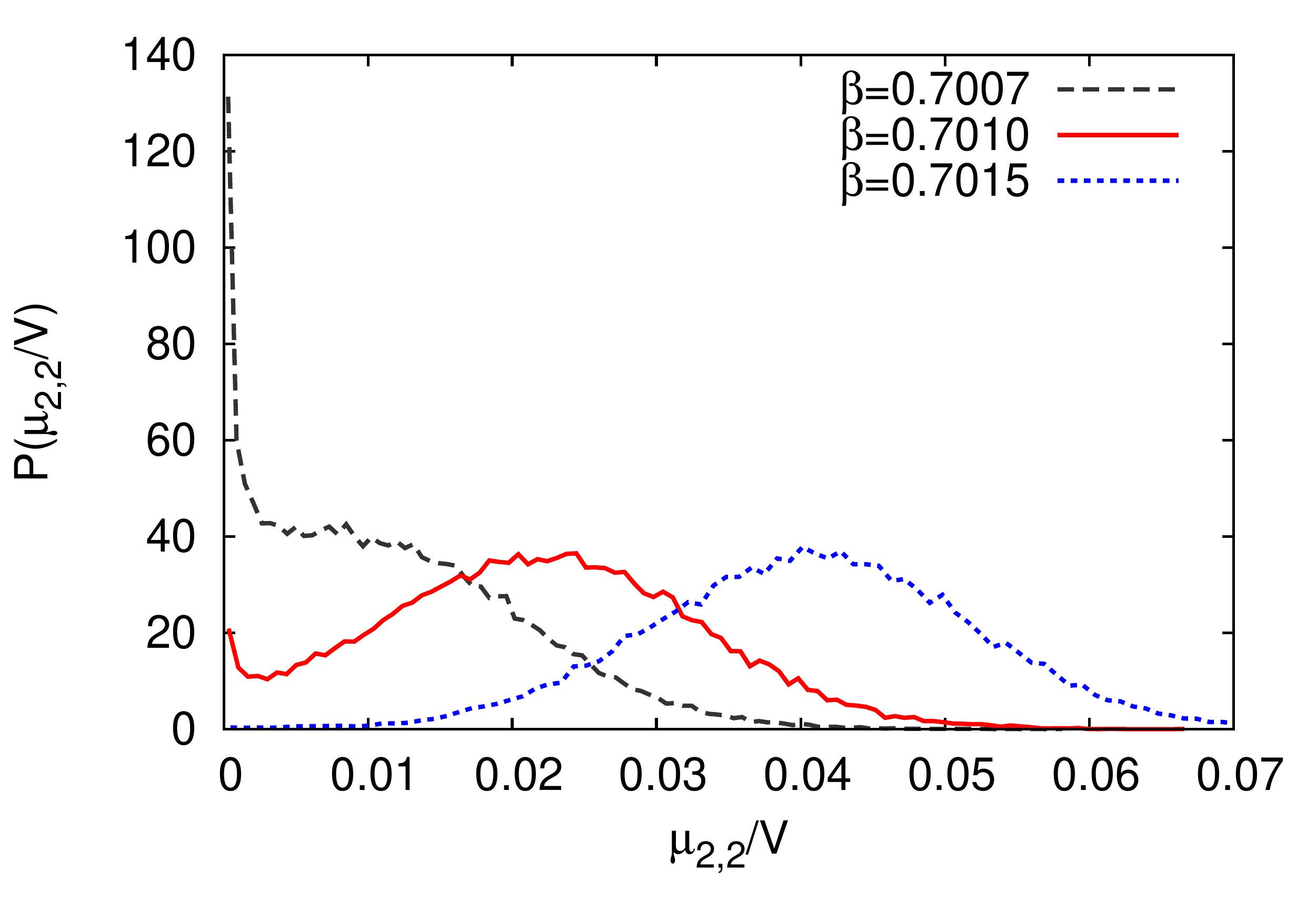}
\caption{Probability distribution of the order parameter $\mu_{2,2}/V$
  ($V = L^3$ is the volume) for the model with $N=2$, $q=4$.  Results
  for $\beta = 0.7007$, $0.7010$, $0.7015$ at fixed $\kappa = 0.4$.
  Here $L = 64$.  }
\label{probmu2-N2q4-kappa0p4}
\end{figure}

To clarify the nature of the transition, we have analyzed $U_2$ and
$R_{\xi,2}$ as a function of $X = (\beta-\beta_c) L^{1/\nu}$. Again,
results are not consistent with an Ising behavior. Indeed, repeating
the combined analysis of $U_N$ and $R_{\xi,N}$, as we did before, we
obtain $\nu = 0.56(1)$ if we consider all data, and $\nu = 0.54(1)$,
if only results with $L\ge 24$ are included.  Additional information
on the critical behavior is provided by the analysis of the specific
heats $C_g$ and $C_k$.  They have a pronounced peak that increases
with $L$, see Fig.~\ref{Ck-N2q4-kappa0p4} for a plot of $C_k$.  The
maximum increases apparently as $L^{0.8}$, much more than in the Ising
model, in which it increases as $L^{\alpha/\nu}$ with
$\alpha/\nu\approx 0.17$. Finally, we compute the distributions of the
order parameter $\mu_{2,2}$ defined in Eq.~(\ref{binderdef}).  In
Fig.~\ref{probmu2-N2q4-kappa0p4} we report the results for three
values of $\beta$ and for lattices of size $L=64$.  For $\beta =
0.7010$ we observe the presence of two maxima, a hint for a
first-order transition: a very sharp one for $\mu_{2,2}\approx 0$ and
a broad one at a finite value of $\mu_{2,2}$.

Collecting all results we conclude that the critical behavior
along the DC-OD transition line for $N=2$ is not the same as for $N\ge
3$.  The most likely possibility is that the Ising transition that
occurs for $\kappa=0$ (for this value of the gauge coupling, the model
can be mapped exactly onto the Ising model, see App.~\ref{App.A})
turns into a first-order transition at some critical value $\kappa^*$:
for $\kappa < \kappa^*$ we have an Ising transition, for $\kappa >
\kappa^*$ the transition becomes of first order, while for $\kappa =
\kappa^*$ there is a tricritical point with mean-field exponents (in
particular, $\nu = 1/2$) with logarithmic corrections.  We are not
able to estimate $\kappa^*$. We can only infer from the data that
$\kappa^*$ should be smaller than, but not very much different from,
$\kappa = 0.4$, the value at which simulations have been performed.
Indeed, the numerical data in Fig.~\ref{URxi-N2q4-kappa0p4} are close
to the Ising universal curve, indicating the presence of strong Ising
crossover effects that can be explained by the presence of a nearby
Ising transition line. Moreover, the estimates of the critical
exponent $\nu$ and of the specific-heat exponent $\alpha/\nu$ are not
far from the values expected for a tricritical point, $\nu = 1/2$ and
$\alpha/\nu = 1$.

The present results can also be used to predict the behavior of two
Ising systems that interact by means of a ${\mathbb Z}_2$ gauge field
(this is the equivalent interpretation of the ${\mathbb Z}_4$ model
with ${\mathbb Z}_2$ gauge invariance, see App.~\ref{App.A}).  The
gauge interaction, if sufficiently strong, is able to drive the system
far from the Ising fixed point, giving rise to a first-order
transition.

\begin{figure}[tbp]
\includegraphics*[scale=\graphicscale,angle=-90]{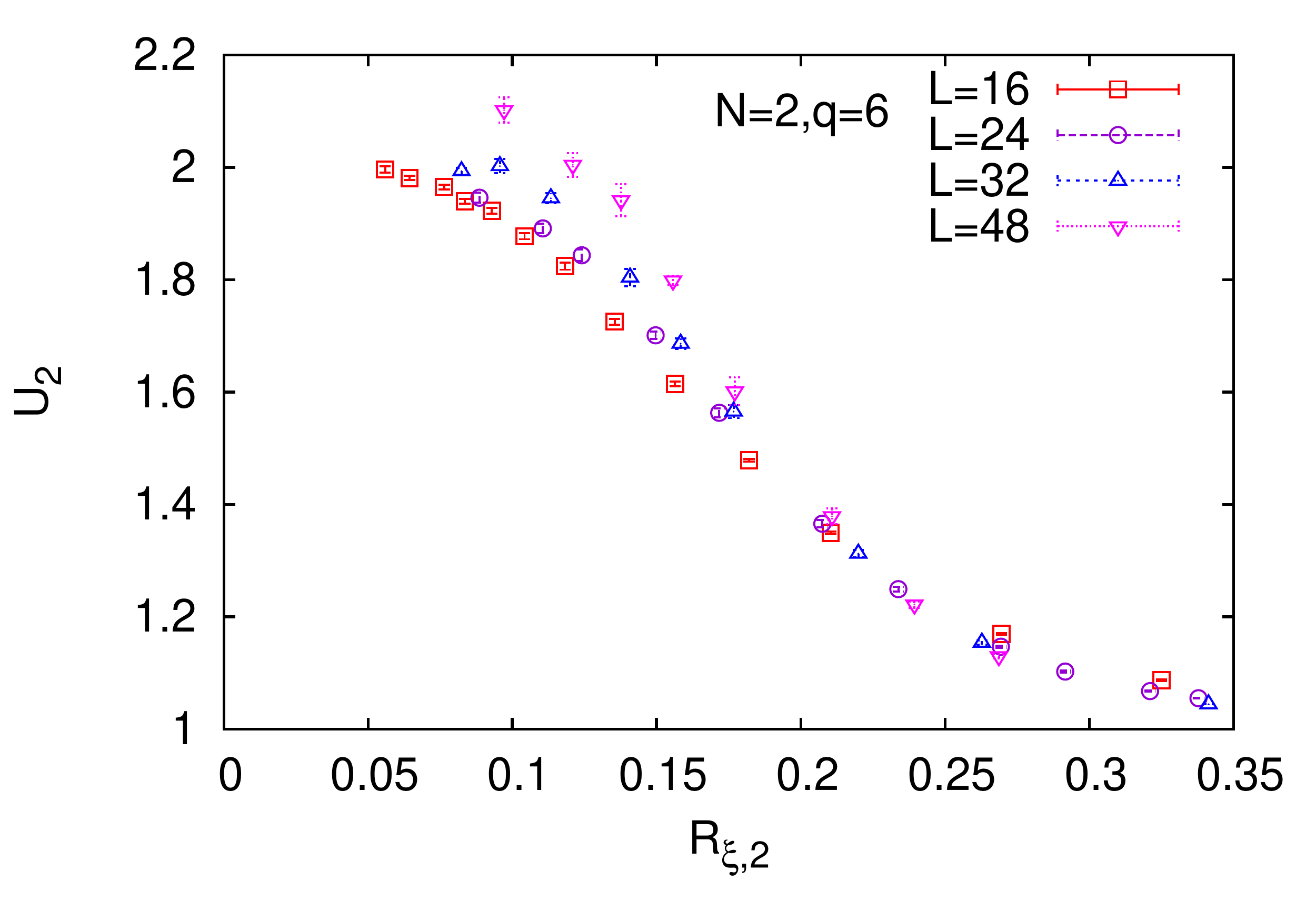}
\includegraphics*[scale=\graphicscale,angle=-90]{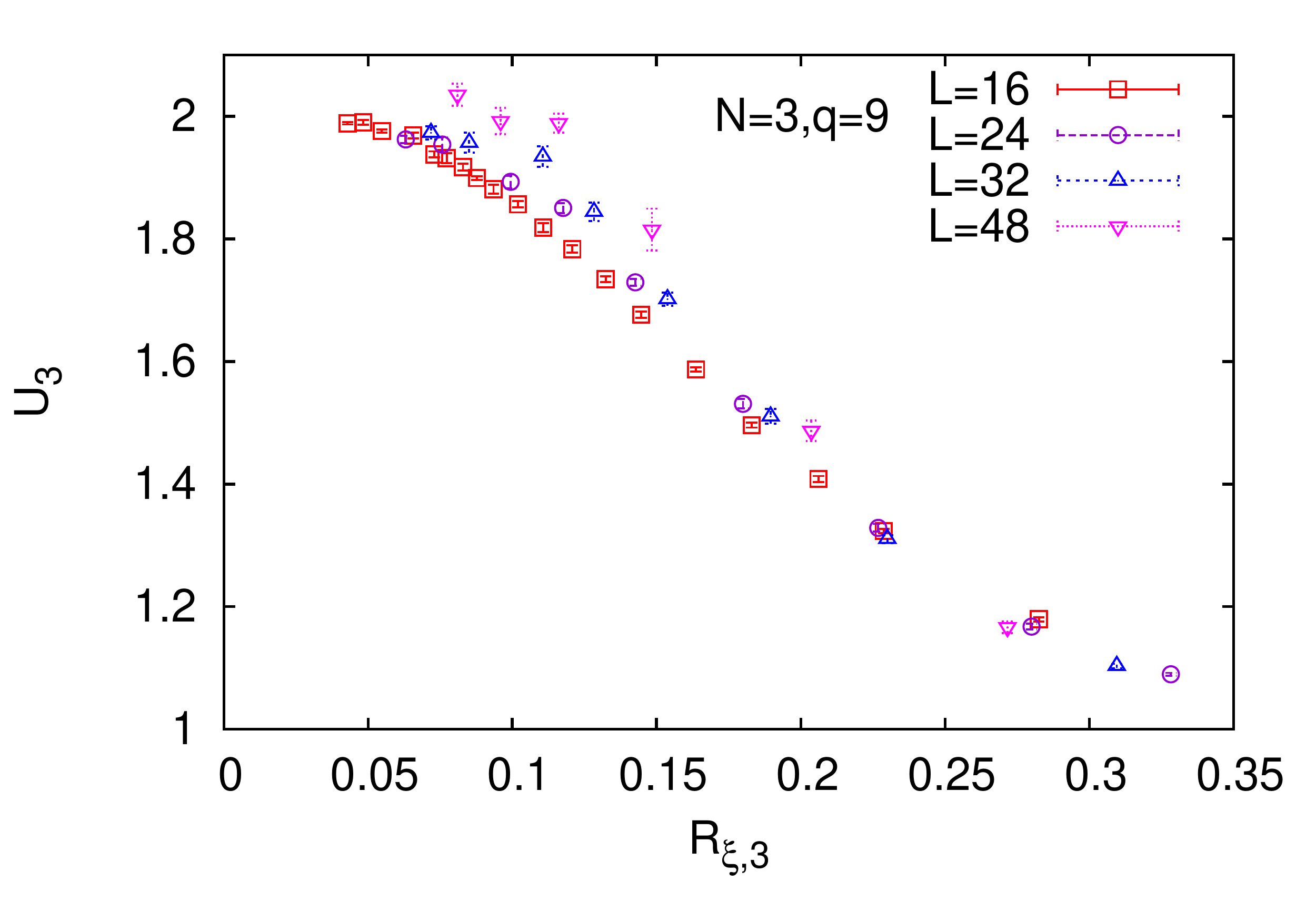}
\caption{Estimates of $U_N$ versus $R_{\xi,N}$ for $N=2$, $q=6$ (top)
  and $N=3$, $q = 9$ (bottom), at fixed $\kappa = 0.4$. In both cases
  $p=q/N=3$.}
\label{URxi-p3-kappa0p4}
\end{figure}

\subsubsection{Models with $p=3$}

For $p = 3$ we have performed simulations for $(q,N) = (6,2)$ and
(9,3).  As expected, in all cases data suggest a first-order
transition, at $\beta_c \approx 0.875 $ and $\beta_c \approx 1.89$,
respectively, as in the ${\mathbb Z}_3$ model. To clarify the nature
of the critical behavior, we have studied the behavior of the Binder
parameters $U_N$ as a function of $R_{\xi,N}$. The results for the two
models are reported in Fig.~\ref{URxi-p3-kappa0p4}. In both cases, we
do not observe scaling. As $L$ increases, the estimates of $U_N$ at
fixed $R_{\xi,N}$ apparently increase, especially for $0.07 \lesssim
R_{\xi,N}\lesssim 0.15$.  In particular, the maximum $U_{\rm max}(L)$
slightly increases as a function of $L$. These results are all
consistent with a first-order transition. It is clear that a
convincing identification of the first-order nature of the transition
requires significantly larger lattices. However, given that these
conclusions are already in agreement with what is expected on the
basis of the arguments of Sec.~\ref{sec4}, we have not further pursued
this issue.

\subsubsection{Models with $p\ge 4$}

\begin{figure}[tbp]
\includegraphics*[scale=\graphicscale,angle=-90]{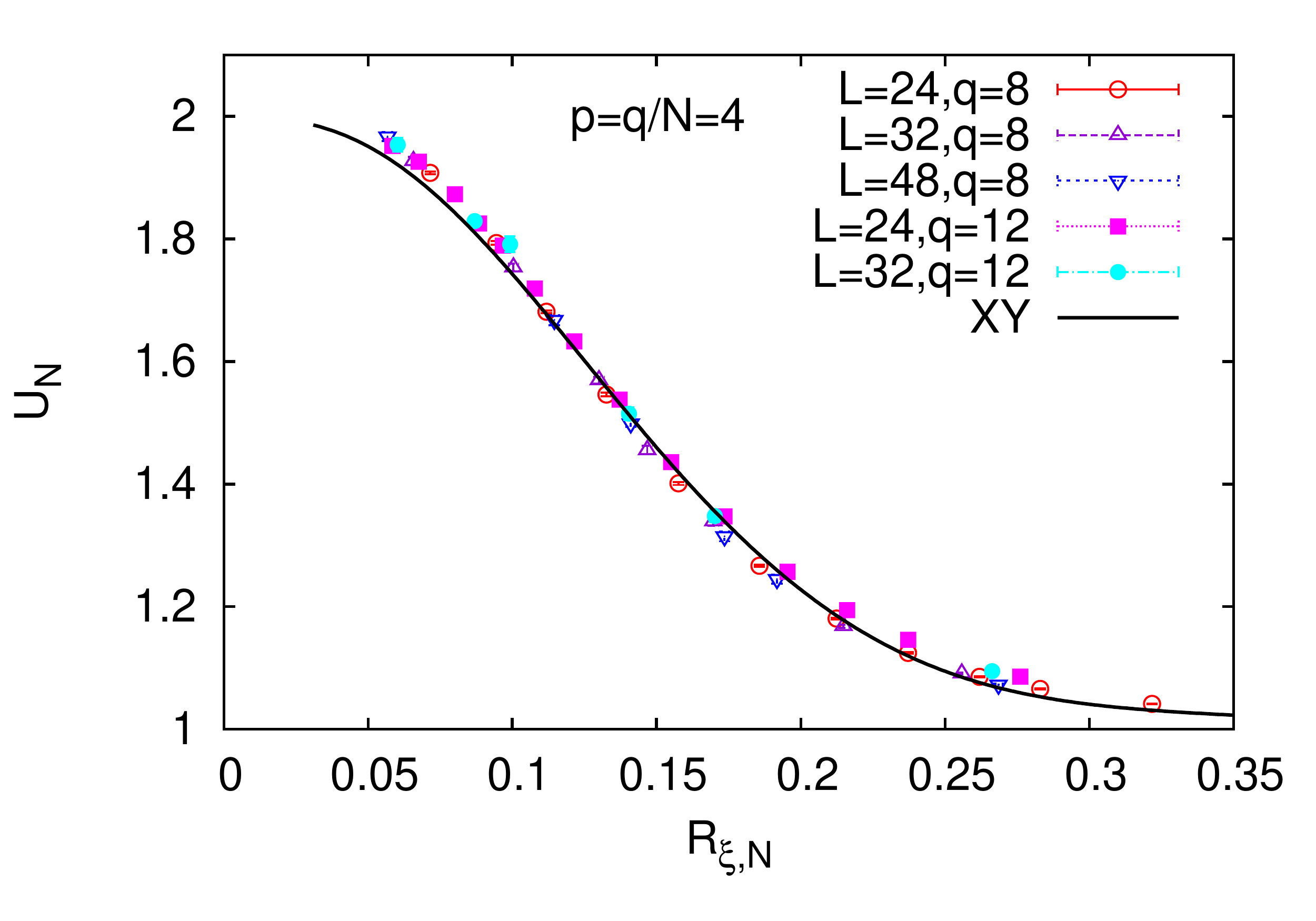}
\includegraphics*[scale=\graphicscale,angle=-90]{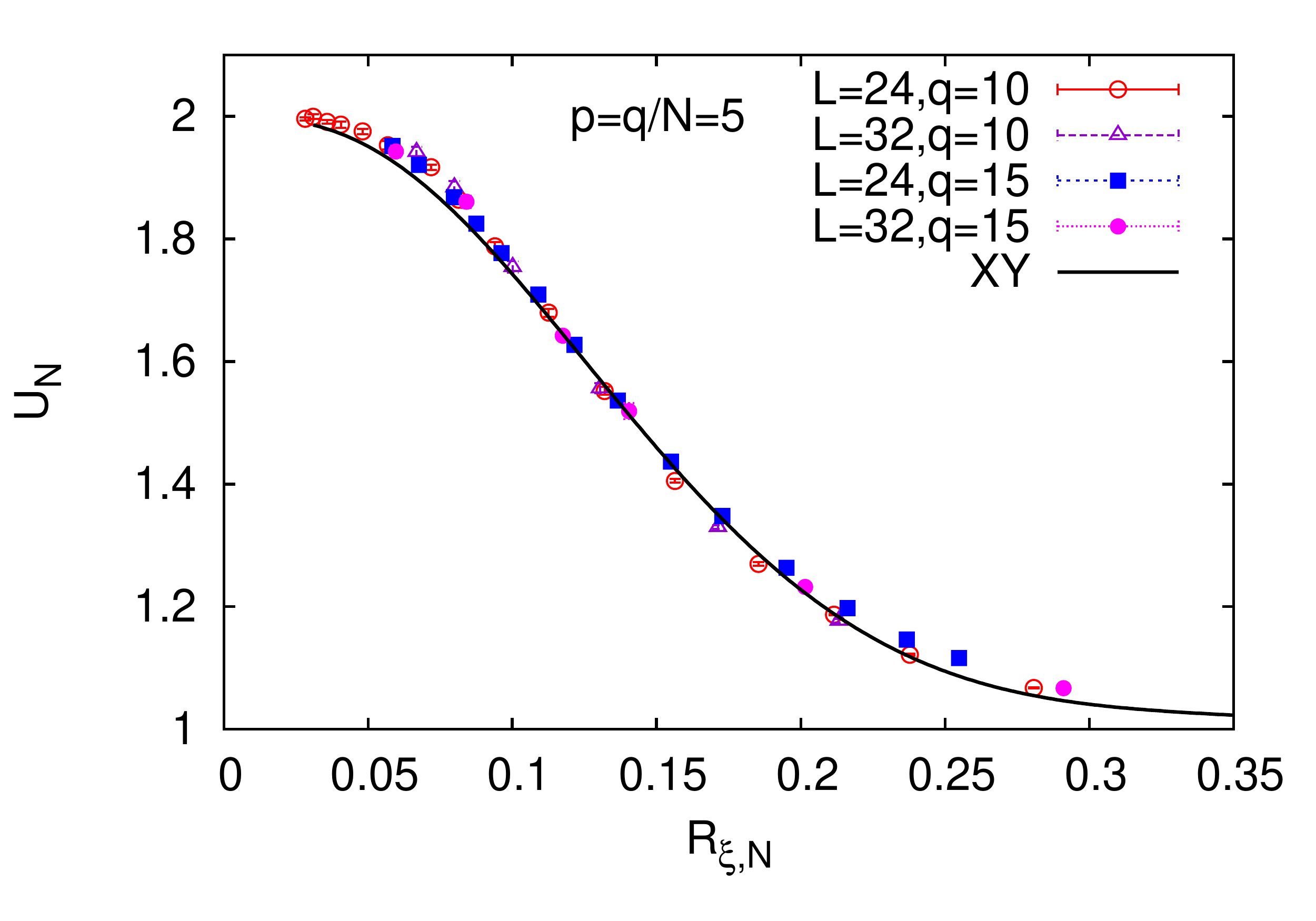}
\caption{Estimates of $U_N$ versus $R_{\xi,N}$ for $N=2$, $q=8$ and
  $N=3$, $q=12$ (data with $p=4$, top) and for $N=2$, $q = 10$ and
  $N=3$, $q=15$ (data with $p=5$, bottom), at fixed $\kappa =
  0.4$. Results are compared with the scaling function appropriate for
  the XY universality class, see App.~\ref{App.C}.  }
\label{URxi-pge4-kappa0p4}
\end{figure}

For $p = 4$ we have performed simulations for $(q,N) = (8,2)$ and
(12,3), while for $p=5$ we have performed simulations for $(q,N) =
(10,2)$ and (15,3).  For both values of $p$, data are consistent with
an XY behavior, see Fig.~\ref{URxi-pge4-kappa0p4}. While for $p=5$
this is the same behavior as observed in the ${\mathbb Z}_5$ model,
for $p=4$ we do not observe the Ising behavior characterizing the
clock ${\mathbb Z}_4$ model. On one side, this is expected, since in
generic models with ${\mathbb Z}_4$ symmetry breaking one expects the
emergence of an enlarged O(2) symmetry. On the other side, however, it
is somewhat surprising to observe such a good agreement, given that we
expect very slowly-decaying corrections (behaving approximately as
$L^{-0.1}$) due to the spin-four operator that breaks the O(2)
symmetry down to ${\mathbb Z}_4$. We have no evidence of such
corrections in the plots of $U_N$ versus $R_{\xi,N}$.

To estimate the critical point $\beta_c$ we have performed fits to
Eq.~(\ref{rsca}). Assuming the transition to belong to the XY
universality class, we have fixed $\nu = 0.6717$ and $\omega = 0.789$
\cite{Hasenbusch-19}. Results show a very tiny dependence on $q$.  For
$N=2$ we obtain $\beta_c = 0.8869(1)$ and 0.8869(2) for $q = 8$ and
10, respectively. For $N=3$, we have $\beta_c = 1.9160(15)$,
1.9150(15) for $q = 12$ and 15.

\subsubsection{Summary and Landau-Ginzburg-Wilson effective theory}

The numerical simulations confirm the predictions of Sec.~\ref{sec4}.
The only relevant variable along the DC-OD line is the ratio $p =
q/N$.  For $p=2$ we confirm that the models belong to the Ising
universality class, with one only exception, the model with $q = 4$
and $N=2$, which undergoes a first-order transition, at least for not
too small values of $\kappa$.  For $p=3$, the transition is apparently
of first order, as in the ${\mathbb Z}_3$ clock model. For $p\ge 4$,
we observe XY behavior in all cases, including $p=4$. Note that in the
latter case the ${\mathbb Z}_4$ clock model has an Ising transition
due to a peculiar factorization of the degrees of freedom, see
App.~\ref{App.A}.

These results have a very simple interpretation in the
Landau-Ginzburg-Wilson (LGW) framework. In this approach, one assumes
that the critical behavior is completely determined by the
gauge-invariant scalar modes, so that it can be determined by studying
the effective Hamiltonian $H_{\rm LGW}$ for a coarse-grained
gauge-invariant order parameter that is invariant under the global
symmetry group of the model.  For the model we consider, the
microscopic order parameter is $w^N$ and the global symmetry group is
${\mathbb Z}_q/{\mathbb Z}_N = {\mathbb Z}_p$. For $p = 2$, $w^N$ is
real and therefore, we must consider a LGW model for a scalar real
field with ${\mathbb Z}_2$ global invariance. Such a model describes
the standard Ising behavior. For $p > 2$, $w^N$ is a complex number,
so that the fundamental field is a complex field $\psi$. The effective
Hamiltonian density is
\begin{equation}
{\cal H}_{\rm LGW} = 
    \left(\sum_\mu \partial_\mu \bar{\psi}\partial_\mu {\psi}\right) + 
    r |{\psi}|^2 + u |{\psi}|^4 + g_p (\psi^p + \bar{\psi}^p) + \ldots
\label{LGW}
\end{equation}
For $p > 4$, the terms with coefficient $g_p$ are irrelevant, and thus
we obtain the O(2)/XY LGW model. For $p=4$, the Hamiltonian
(\ref{LGW}) is equivalent to that of the so-called cubic
model~\cite{CPV-00} for a two-component real field. A RG analysis
shows that continuous transitions in this class of models belong to
the XY universality class: the cubic-symmetric perturbation
proportional to $g_p$ is irrelevant~\cite{CPV-00} at the XY fixed
point.  For $p=3$, the approach predicts a first-order transition
because of the presence of a cubic term.

\subsection{Large-$\kappa$ transition line }

\begin{figure}[tbp]
\includegraphics*[scale=\graphicscale,angle=-90]{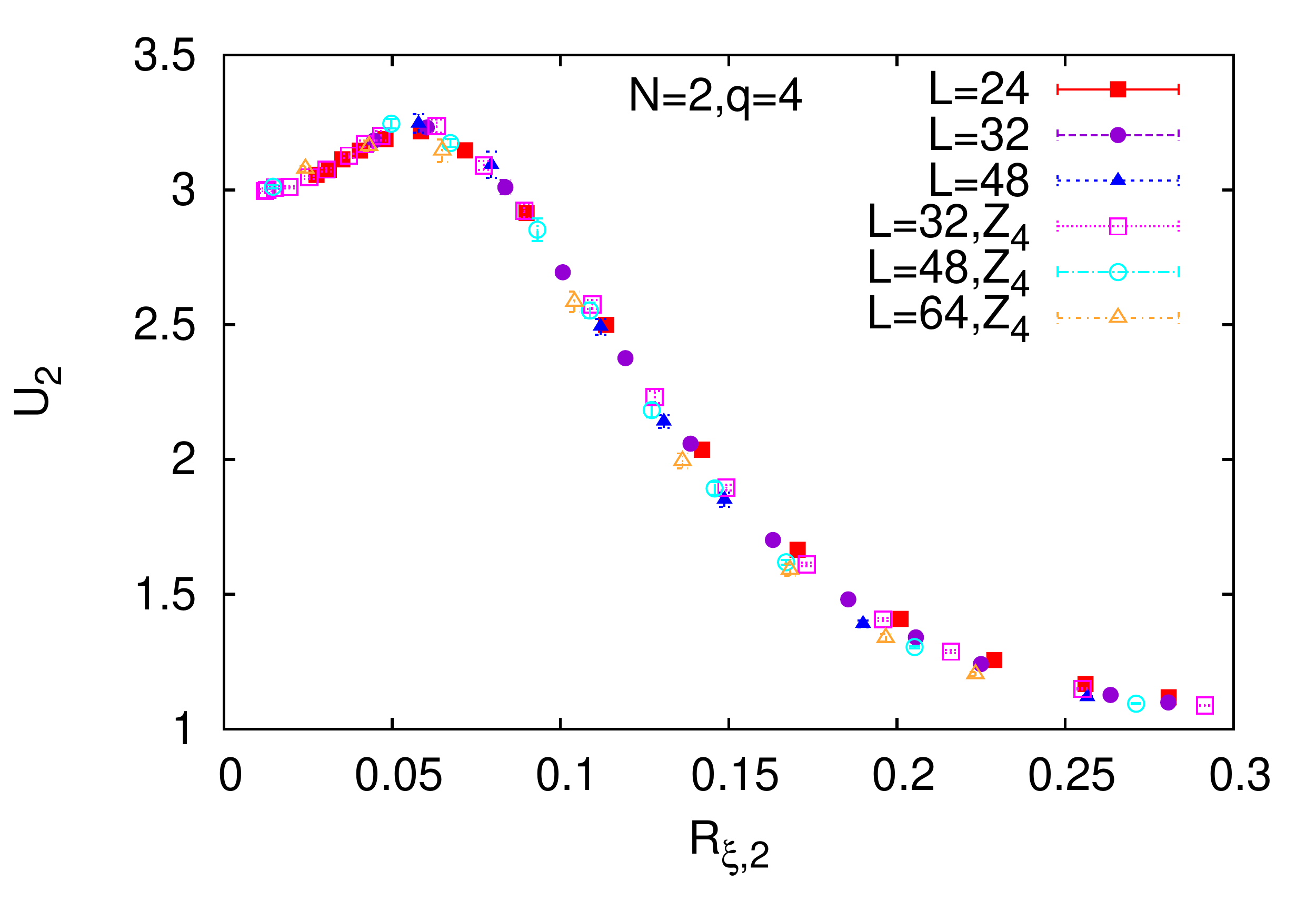}
\caption{Estimates of $U_2$ versus $R_{\xi,2}$ along the DD-OD line
  for $q=4$.  Results for $N=2$ along the line $\kappa = 1$.  We also
  report results for ${\mathbb Z}_4$ clock model.  }
\label{URxi-N2q4-kappalarge}
\end{figure}

\begin{figure}[tbp]
\includegraphics*[scale=\graphicscale,angle=-90]{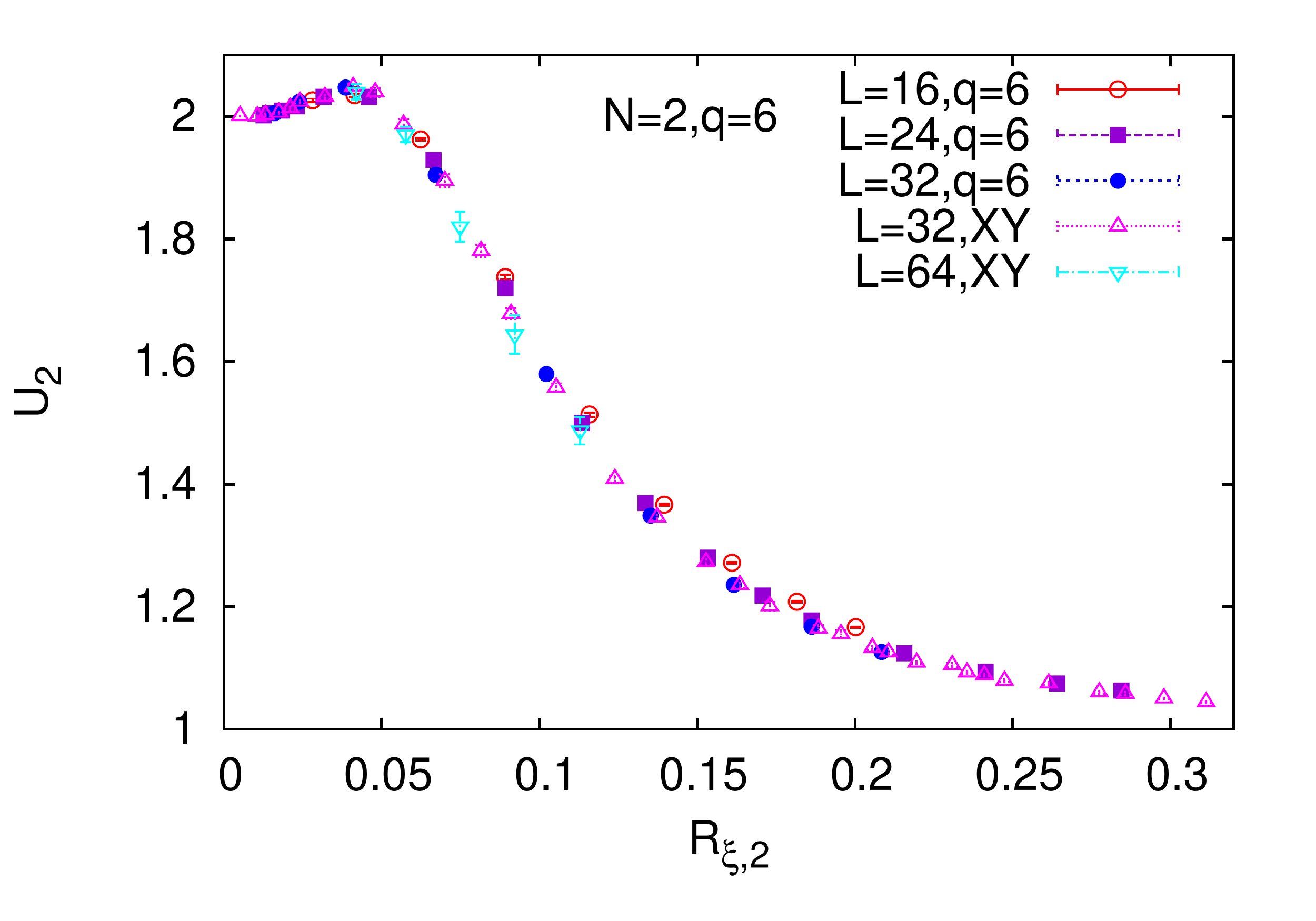}
\includegraphics*[scale=\graphicscale,angle=-90]{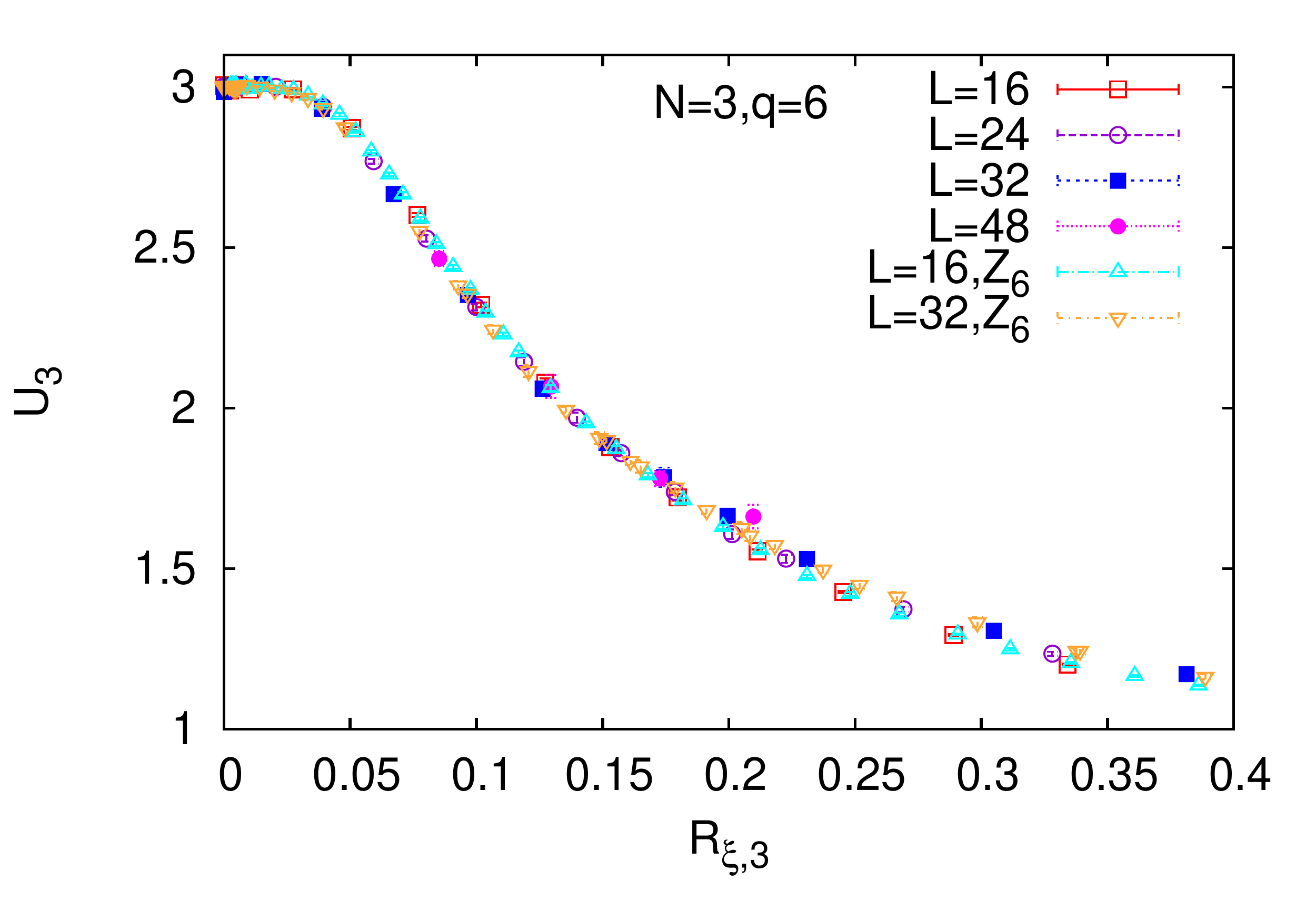}
\caption{Estimates of $U_N$ versus $R_{\xi,N}$ along the DD-OD line,
  for $q=6$.  Results for $N=2$ along the line $\kappa = 1$ (top), and
  for $N=3$ along the line $\kappa = 1.5$ (bottom). In the upper panel
  we also report results for the XY model, in the lower panel for the
  ${\mathbb Z}_6$ clock model.  }
\label{URxi-kappalarge}
\end{figure}

We have studied the behavior along the DD-OD line for two values of
$N$, $N=2$ and $N=3$. We have fixed $\kappa = 1$ and $\kappa = 1.5$ in
the two cases, respectively. These two values have been chosen on the
basis of the numerical estimates for the location of the topological
DC-DD transition line.  For $N=2$, we have a transition at $\kappa_c
\approx 0.76$, both for $\beta = 0$ (see Sec.~\ref{sec4}) and for
$\beta = 0.2$. For $N=3$, we have a transition at $\kappa_c \approx
1.08$ both for $\beta = 0$ \cite{BCCGPS-14} and for $\beta = 0.2$. In
both cases, along the DC-DD transition line, $\kappa$ is essentially
constant.  This guarantees us that, for the two chosen values of
$\kappa$, we are considering transitions along the DC-DD line.

We have performed simulations for $(q,N) = (4,2)$, (6,2), and (6,3)
observing an ordering transition at $\beta_c = 0.4437(1)$, 0.4541(3),
0.4555(10), respectively. To further check that the transition belongs
to the DD-OD line we have determined the gauge energy $E_g$. Close to
the transition we find $E_g\approx 2.98$, 2.99 for $N=2$ and 3,
respectively: most of the plaquettes are indeed equal to 1 (for
$\kappa \to \infty$ we have $E_g = 3$).

As we have discussed in Sec.~\ref{sec4}, we expect the model to behave
as the ungauged ${\mathbb Z}_q$ model. Our results are in full
agreement.  In Fig.~\ref{URxi-N2q4-kappalarge} we report the results
for $U_2$ versus $R_{\xi,2}$ for $q=4$ and compare them with the
analogous results for the ${\mathbb Z}_4$ model.  We observe a very
good agreement. Clearly, the presence of the gauge interaction is
unable to destabilize the decoupled Ising behavior, as it does along
the DC-OD transition line. In Fig.~\ref{URxi-kappalarge} we report
results for the Binder parameters for $q=6$.  Since the transition in
the ${\mathbb Z}_6$ clock model is in the universality class of the XY
transition, one might think of comparing the scaling curves with those
computed in the XY model.  However, this is only possible for $U_1$
and $U_2$, but not for $U_3$, as discussed in
App.~\ref{App.C}. Therefore, the results for $U_3$ obtained in the
gauge-scalar model are directly compared with the ${\mathbb Z}_6$
results.  In all cases, we observe very good agreement, confirming the
irrelevance of the gauge coupling along the DD-OD line

\section{Conclusions} \label{sec6}

In this work we have studied a gauge-Higgs model with discrete scalar
fields and ${\mathbb Z}_N$ gauge invariance. It is obtained by gauging
the ${\mathbb Z}_N$ subgroup of the global invariance group of the
${\mathbb Z}_q$ clock model ($N$ is a submultiple of $q$), in which
the scalar fields are phases that take the $q$ values $\exp(2 \pi i
n/q)$, $n=0,\ldots q-1$. The resulting model is invariant under local
${\mathbb Z}_N$ and global ${\mathbb Z}_q/{\mathbb Z}_N = {\mathbb
  Z}_p$ ($p=q/N$) transformations. The phase diagram of the model is
reported in Fig.~\ref{phdia}. There are three different transition
lines. On one line one expects the topological transitions that
characterize the pure gauge ${\mathbb Z}_N$ theory. We have studied in
the detail the behavior of the model along the other two transition
lines, along which the scalar field orders.

The critical behavior along the small-$\kappa$ transition line that
separates the disordered-confined phase from the ordered-deconfined
phase turns out to be in full agreement with the predictions of the
Landau-Ginzburg-Wilson approach. Criticality depends only on the
global ${\mathbb Z}_p$ symmetry group of the effective theory, so that
the model behaves as a ferromagnetic system with a one component
complex field and ${\mathbb Z}_p$ global invariance. We thus predict
that continuous transitions belong to the Ising universality class for
$p=2$ and to the O(2) universality class for any $p \ge 4$, the
${\mathbb Z}_p$ breaking terms being dangerously irrelevant
perturbations.  For $p=3$ instead only first-order transitions are
possible.  Numerical data confirm these predictions quite
precisely. In particular, we verify that symmetry enlargement occurs
at the transition, as in the standard clock model. The condition $q\ge
4$ is now replaced by $p = q/N\ge 4$, consistently with the idea that
$p$ counts the effective number of degrees of freedom per site.

For large $\kappa$, the OD-DD line separates two phases, in which the
gauge fields are deconfined, see Fig.~\ref{phdia}. On this line, gauge
fields do not play any role (modulo gauge transformations, we have
$\sigma_{x,\mu} = 1$ on most of the links), and the model behaves as
the ${\mathbb Z}_q$ clock model, irrespective of the values of $N$.

\appendix

\section{Relation between the ${\mathbb Z}_4$ model with 
${\mathbb Z}_2$ gauge invariance and the Ising model} \label{App.A}

We wish to relate the ${\mathbb Z}_4$ model with ${\mathbb Z}_2$ gauge
invariance with an Ising system. We rewrite $w_{\bm x} =
e^{i\theta_{\bm x}}$ and parametrize the field in terms of two Ising
spins $\tau^{(1)}_{\bm x}$ and $\tau^{(2)}_{\bm x}$, as
\begin{equation}
\cos\theta_{\bm x} = {1\over2} (\tau^{(1)}_{\bm x} + \tau^{(2)}_{\bm
  x})\,, \quad \sin\theta_{\bm x} = {1\over2} (\tau^{(1)}_{\bm x} -
\tau^{(2)}_{\bm x}) \,.
\end{equation}
In terms of the Ising spins, the Hamiltonian $H_{\rm kin}$ becomes
\begin{equation}
{1\over T} H_{\rm kin} = - {\beta\over2} \sum_{{\bm x}\mu} 
   (\tau^{(1)}_{\bm x} \tau^{(1)}_{{\bm x} + \hat{\mu}} + 
    \tau^{(2)}_{\bm x} \tau^{(2)}_{{\bm x} + \hat{\mu}} ) 
    \sigma_{{\bm x},\hat{\mu}}\,,
\end{equation}
which shows that the model is equivalent to two Ising systems coupled
by the gauge field. Correspondingly, the correlation functions
$G_{Q}({\bm x},{\bm y})$ become
\begin{eqnarray}
G_1({\bm x},{\bm y}) &=& {1\over 2} \langle \tau^{(1)}_{\bm x}
\tau^{(1)}_{\bm y} + \tau^{(2)}_{\bm x} \tau^{(2)}_{\bm y} \rangle\,,
\nonumber \\ G_2({\bm x},{\bm y}) &=& \langle \tau^{(1)}_{\bm x}
\tau^{(2)}_{\bm x} \tau^{(1)}_{\bm y} \tau^{(2)}_{\bm y} \rangle \,.
\end{eqnarray}
In the absence of the gauge fields, i.e., in the ${\mathbb Z}_4$ clock
model, the two Ising models decouple, and $G_1({\bm x},{\bm y})$
corresponds to the two-point function in the Ising model. In the same
limit, the Binder parameter $U_1$ satisfies
\begin{equation}
   U_1 = {1\over 2} U_{\rm Is} + {1\over 2} \,,
\label{binderZ4_Q1}
\end{equation}
where $U_{\rm Is}$ is the Binder parameter in the Ising model. 

For $\kappa\to 0$, the gauge fields can be integrated out. If $a$ and $b$ 
can only take the values $\pm 1$, we can easily prove the identity
\begin{eqnarray}
&& \sum_{\sigma=\pm 1} e^{K a \sigma}\,e^{K b \sigma} =
  2 \cosh^2 K + 2 a b \sinh^2 K\, .
\end{eqnarray}
If we define 
$\tilde{\beta}$ and $A$ as 
\begin{equation}
\tanh \tilde{\beta} = \tanh^2 {\beta\over 2}\,, \quad
A = 2 \left( \cosh^2 {\beta\over2} + \sinh^2 {\beta\over2} \right)^{1/2}\,, 
\end{equation}
and a new Ising spin $\rho_{\bm x}= \tau^{(1)}_{\bm x} \tau^{(2)}_{\bm x}$,
we can rewrite the partition function as 
\begin{equation}
Z = A^{3L} 2^L \sum_{\rho_{\bm x}} e^{-\tilde{\beta}\, H_{\rm eff}}\,,
\quad 
H_{\rm eff} = - \sum_{{\bm x},\mu}
     \rho_{\bm x} \rho_{{\bm x} + \hat{\mu}}\,. 
\end{equation}
We have thus obtained an Ising model for a single spin variable at
inverse temperature $\tilde{\beta}$, The mapping allows us to compute
the critical temperature.  Using $\tilde{\beta}_c = 0.221654626(5)$
\cite{FXL-18}, we obtain $\beta_c = 1.01246856(1)$.

\section{The ${\mathbb Z}_8$ model with 
${\mathbb Z}_2$ gauge invariance} \label{App.B}

One can generalize the considerations of App.~\ref{App.A} to the
${\mathbb Z}_8$ clock model. In this case the field $w_{\bm x} =
e^{i\theta_{\bm x}}$ can be parametrized in terms of three Ising spins
$\tau^{(i)}_{\bm x}$, $i=1,2,3$, so that
\begin{eqnarray}
\cos\theta_{\bm x} &=& 
  {1\over8} (\tau^{(1)}_{\bm x} + \tau^{(2)}_{\bm x}) 
   [2 + \sqrt{2} - ( 2 - \sqrt{2} ) \tau_{\bm x}^{(3)}] \nonumber \\
  && \quad -
  {\sqrt{2}\over 8} (\tau^{(1)}_{\bm x} - \tau^{(2)}_{\bm x}) 
   (1 + \tau_{\bm x}^{(3)})\,, \\
\sin\theta_{\bm x} &=& 
  {1\over8} (\tau^{(1)}_{\bm x} - \tau^{(2)}_{\bm x}) 
   [2 + \sqrt{2} - ( 2 - \sqrt{2} ) \tau_{\bm x}^{(3)}] \nonumber  \\
  && \quad +
  {\sqrt{2}\over 8} (\tau^{(1)}_{\bm x} + \tau^{(2)}_{\bm x}) 
   (1 + \tau_{\bm x}^{(3)})\,. \nonumber 
\end{eqnarray}
Under a gauge transformation $w_{\bm x} \to -w_{\bm x}$,
 the three Ising spins transform as 
\begin{equation}
\tau^{(1)}_{\bm x} \to - \tau^{(1)}_{\bm x}\,, \quad
\tau^{(2)}_{\bm x} \to - \tau^{(2)}_{\bm x}\,, \quad
\tau^{(3)}_{\bm x} \to \tau^{(3)}_{\bm x}\,. \quad
\end{equation}
Using this parametrization, we can rewrite
\begin{eqnarray} 
&& \cos (\theta_{\bm x} - \theta_{\bm y}) = 
   (\tau^{(1)}_{\bm x} \tau^{(1)}_{\bm y} + 
    \tau^{(2)}_{\bm x} \tau^{(2)}_{\bm y}) 
   (a + b \tau^{(3)}_{\bm x} \tau^{(3)}_{\bm y}) \nonumber \\
&& \qquad 
   + c (\tau^{(3)}_{\bm x} - \tau^{(3)}_{\bm y}) 
    (\tau^{(1)}_{\bm y} \tau^{(2)}_{\bm x} - 
     \tau^{(1)}_{\bm x} \tau^{(2)}_{\bm y})\,, 
\end{eqnarray}
where
\begin{equation}
a = {1\over 4\sqrt{2}}(\sqrt{2} + 1)\,, \quad 
b = {1\over 4\sqrt{2}}(\sqrt{2} - 1)\,, \quad 
c = {1\over 4 \sqrt{2}}\,.
\end{equation}
Let us now consider the model in the presence of a ${\mathbb Z}_2$
gauge field $\sigma$. At $\kappa = 0$ we can integrate out the gauge
field. We compute
\begin{equation}
Z_{{\bm x},{\bm y}} = \sum_{\sigma=\pm 1} 
   e^{\beta \sigma \cos (\theta_{\bm x} - \theta_{\bm y})}\,,
\end{equation}
obtaining
\begin{eqnarray} 
Z_{{\bm x},{\bm y}} &=& 2 \cosh^2 {\beta a} \cosh^2 {\beta b}
      \cosh^4 {\beta c} \left(1 + A_1 +  \right. \nonumber \\
  &&   A_2 \tau^{(3)}_{\bm x} \tau^{(3)}_{\bm y} + 
     A_3 \tau^{(1)}_{\bm x} \tau^{(1)}_{\bm y}
         \tau^{(2)}_{\bm x} \tau^{(2)}_{\bm y} + \nonumber \\
  &&    \left. 
   A_4 \tau^{(1)}_{\bm x} \tau^{(1)}_{\bm y}
         \tau^{(2)}_{\bm x} \tau^{(2)}_{\bm y} 
         \tau^{(3)}_{\bm x} \tau^{(3)}_{\bm y} \right)\,,
\label{ZZ8}
\end{eqnarray}
where 
\begin{eqnarray}
 A_1 &=& - 2(t_a^2 + t_b^2) t_c^2 + t_a^2 t_b^2 (1 + t_c^4) + t_c^4 \,,
\nonumber \\
 A_2 &=& 2 t_a t_b (1 - t_c^2)^2 - 2 t_c^2 (1 - t_a^2) (1 - t_b^2) \,,
\nonumber \\
 A_3 &=& (t_a^2 + t_b^2) (1 + t_c^4) - 2 t_c^2 (1 + t_a^2 t_b^2) \,,
\nonumber \\
 A_4 &=& 2 t_a t_b (1 - t_c^2)^2 + 2 t_c^2 (1 - t_a^2) (1 - t_b^2)\,,
\end{eqnarray}
and we have defined $t_a = \tanh \beta a$, $\ldots$ An important
property of the result is the relation $A_3=A_4$, which is not
apparent from the previous expressions. To prove it, it is necessary
to express $t_a$ and $t_b$ in terms of $t_c$ and $t_d$, where $d =
1/4$.  We end up with
\begin{equation}
A_3 = A_4 = {2 (1 - t_c^2)^4 t_d^2\over (1 - t_c^2 t_d^2)^2}\,.
\end{equation}
Eq.~(\ref{ZZ8}) shows that the model can be parametrized in terms 
of two Ising fields. We define 
\begin{equation}
    \rho^{(1)}_{\bm x} = 
      \tau^{(1)}_{\bm x} \tau^{(2)}_{\bm x} \tau^{(3)}_{\bm x}\,, 
\qquad 
    \rho^{(2)}_{\bm x} = \tau^{(1)}_{\bm x} \tau^{(2)}_{\bm x}\,,
\end{equation}
obtaining the relation
\begin{eqnarray} 
Z_{{\bm x},{\bm y}} &=& K [1 + A_1 + 
       A_2 \rho^{(1)}_{\bm x} \rho^{(1)}_{\bm y}
           \rho^{(2)}_{\bm x} \rho^{(2)}_{\bm y} + \nonumber \\
  &&   A_3 (\rho^{(1)}_{\bm x} \rho^{(1)}_{\bm y} + 
            \rho^{(2)}_{\bm x} \rho^{(2)}_{\bm y})]\,, 
\label{ZZ8bis}
\end{eqnarray}
where $K$ is a constant. Since $A_3^2 \not= A_2 (1 + A_1)$, we obtain
two Ising models interacting by means of an energy-energy
term. Moreover, the model is symmetric under the exchange of the two
fields.  In terms of the fields $\rho^{(i)}$, the $Q=2$ correlation
function takes a very simple form:
\begin{equation}
G_2({\bm x},{\bm y}) = {1\over 2} \langle \rho^{(1)}_{\bm x}
\rho^{(1)}_{\bm y} + \rho^{(2)}_{\bm x} \rho^{(2)}_{\bm y} \rangle\,.
\end{equation}
The critical behavior of model (\ref{ZZ8bis}) is well known
\cite{CPV-00}.  The decoupled Ising fixed point (the one that controls
the behavior of the ${\mathbb Z}_4$ clock model) is unstable. If the
transition is continuous, it is controlled by the XY fixed point.

\section{Ising and XY scaling functions for open boundary conditions} 
\label{App.C}

\begin{figure}[tbp]
\includegraphics*[scale=\graphicscale,angle=-90]{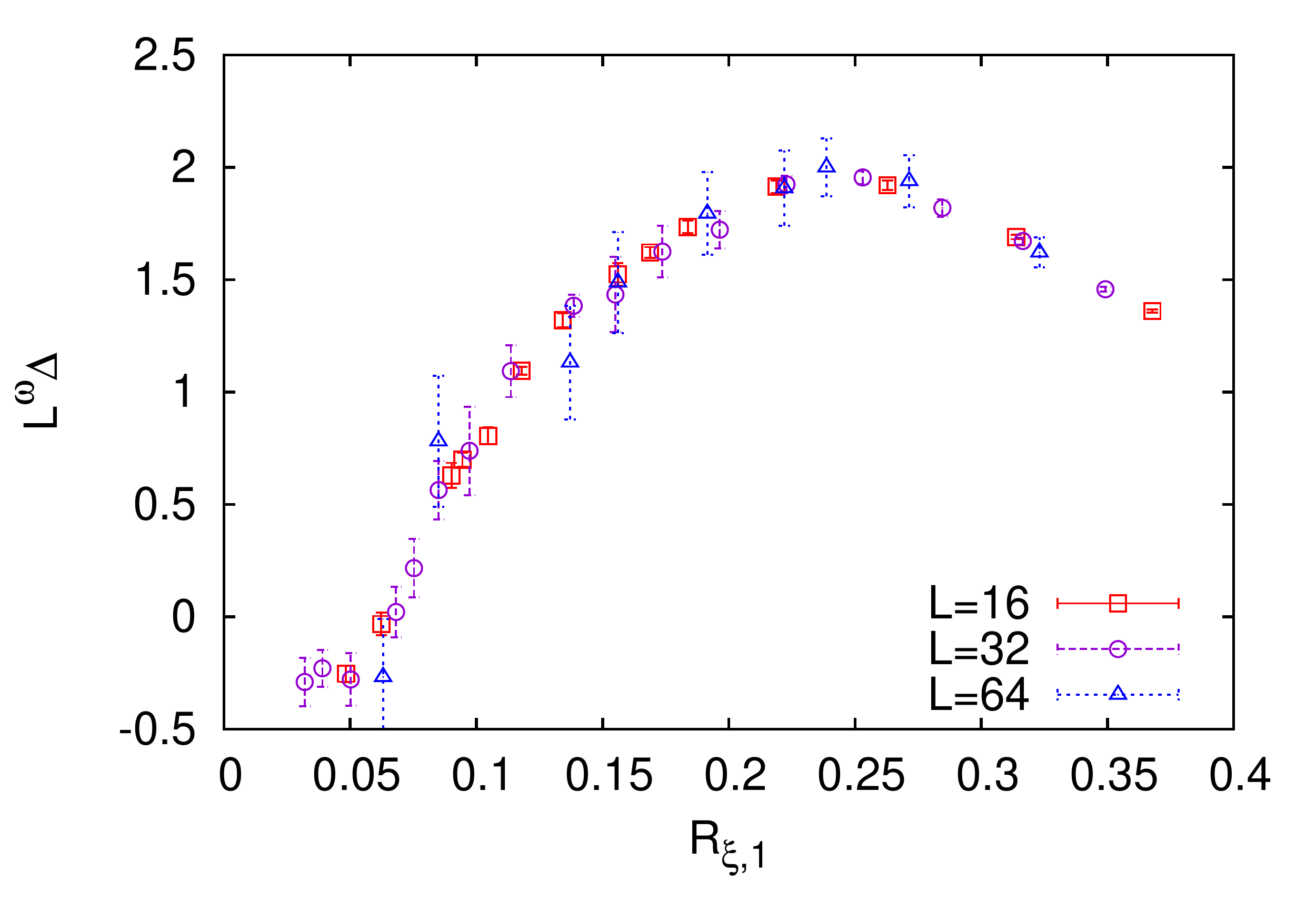}
\includegraphics*[scale=\graphicscale,angle=-90]{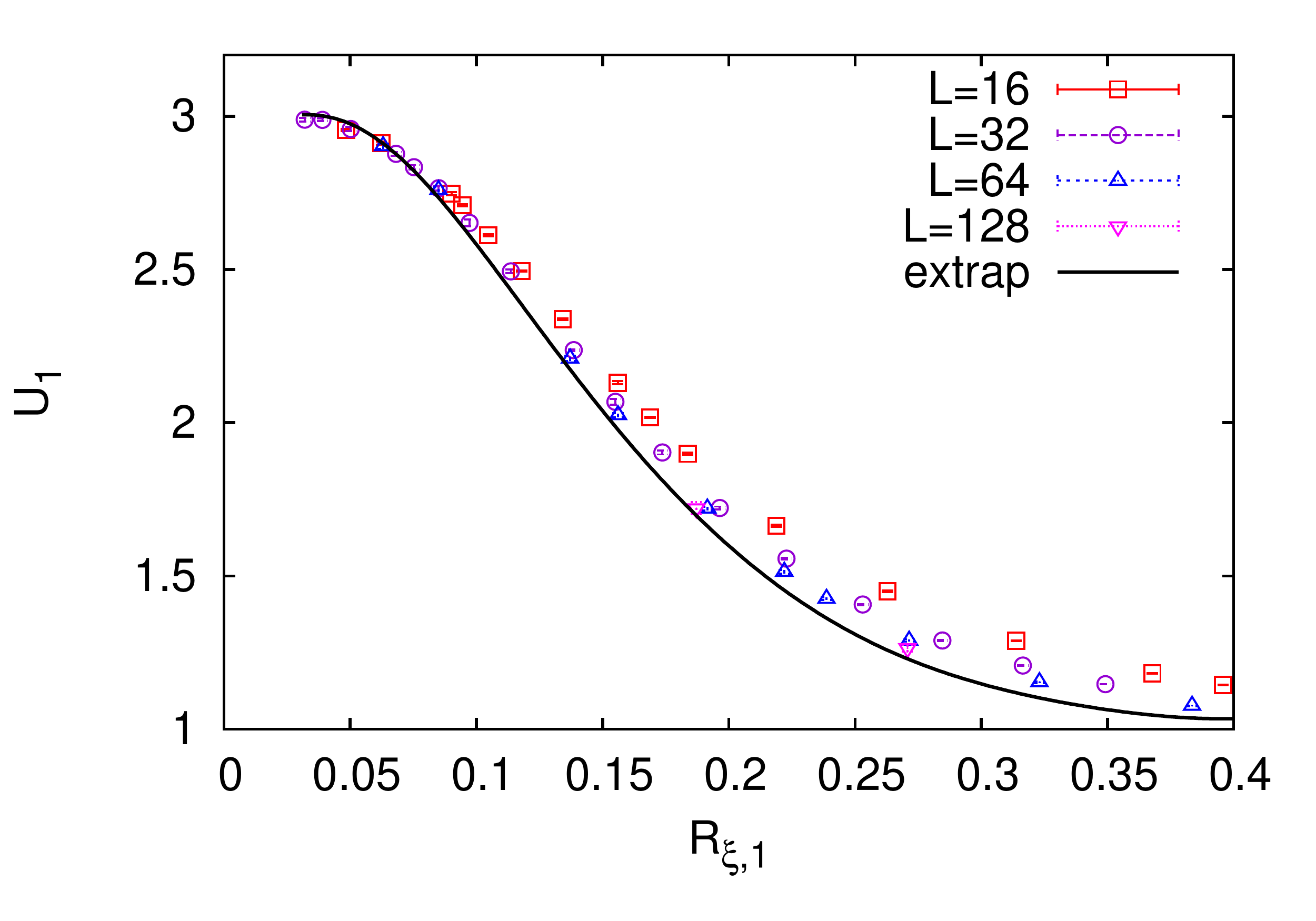}
\caption{Bottom: Data of $U_1$ versus $R_{\xi,1}$ for the Ising model
  and extrapolated scaling function $F_{\xi,1}(R_{\xi,1})$
  (``extrap").  Top: Plot of $L^{\omega}\Delta$ versus $R_{\xi,1}$,
  where $\Delta$ is defined in Eq.~(\ref{deviations-def}) and $\omega
  = 0.83$.  }
\label{URxi-Ising}
\end{figure}

As we have discussed in the text, depending on the values of $N$ and
$q$, we expect to observe either Ising or XY behavior in all cases in
which the transition is continuous. To determine the Ising scaling
curve for $U_1$ as a function of $R_{\xi,1}$ we have performed runs on
lattices of size $L=16,32,64,128$, determining $U_1$ and $\xi_1$. The
results are shown in the lower panel of Fig.~\ref{URxi-Ising}. Scaling
corrections are significant and thus, to obtain the asymptotic scaling
function $F_{\xi,1}(R_{\xi,1})$, we need to perform a fit including
scaling corrections.  We fit the data to Eq.~(\ref{r12sca}),
parametrizing the functions $F(x)$ and $F_c(x)$ with polynomials in
$x$.  We fix $\omega = 0.83$, which is the value predicted for the
Ising universality class in Refs.~\cite{EPPSSV-12-14,KPSV-16}.  The
resulting curve is reported in the lower panel of
Fig.~\ref{URxi-Ising}. To verify the quality of the result, we have
considered the deviations $\Delta$ defined in
Eq.~(\ref{deviations-def}). In the upper panel of Fig.~\ref{URxi-Ising},
we report $L^\omega \Delta$. The data scale nicely on a single curve
with good precision, confirming that our estimate of the asymptotic
scaling function $F_{\xi,1}(R_{\xi,1})$ is reliable and providing us
with an estimate of the correction-to-scaling function
$F_{c,\xi,1}(R_{\xi,1})$. Note that this function is universal apart
from a multiplicative rescaling, and thus we expect scaling
corrections to increase monotonically up to $R_{\xi,1} \approx 0.25$
in all models that belong to the Ising universality class. We have
also determined the value of $R_{\xi,1}$ and $U_1$ at the critical
point, by performing combined fits of the two quantities to
\begin{equation}
  R = f(X) + L^{-\omega} f_c(X)\,,
  \quad X = (\beta - \beta_c) L^{1/\nu}\,,
\end{equation}
using $\beta_c = 0.221654626(5)$ \cite{FXL-18} and $\nu = 0.629971(4)$
\cite{KPSV-16}. We obtain 
\begin{equation}
   U^*_1 = 2.72(2)\,, \qquad R^*_{\xi,1} = 0.086(1)\,.
\end{equation}
These results apply to cubic-symmetric lattices with open boundary
conditions.  They are significantly different from those for periodic
boundary conditions: in this case, for instance, $U^*_1 =
1.60356(15)$~\cite{FXL-18}.

\begin{figure}[tbp]
\includegraphics*[scale=\graphicscale,angle=-90]{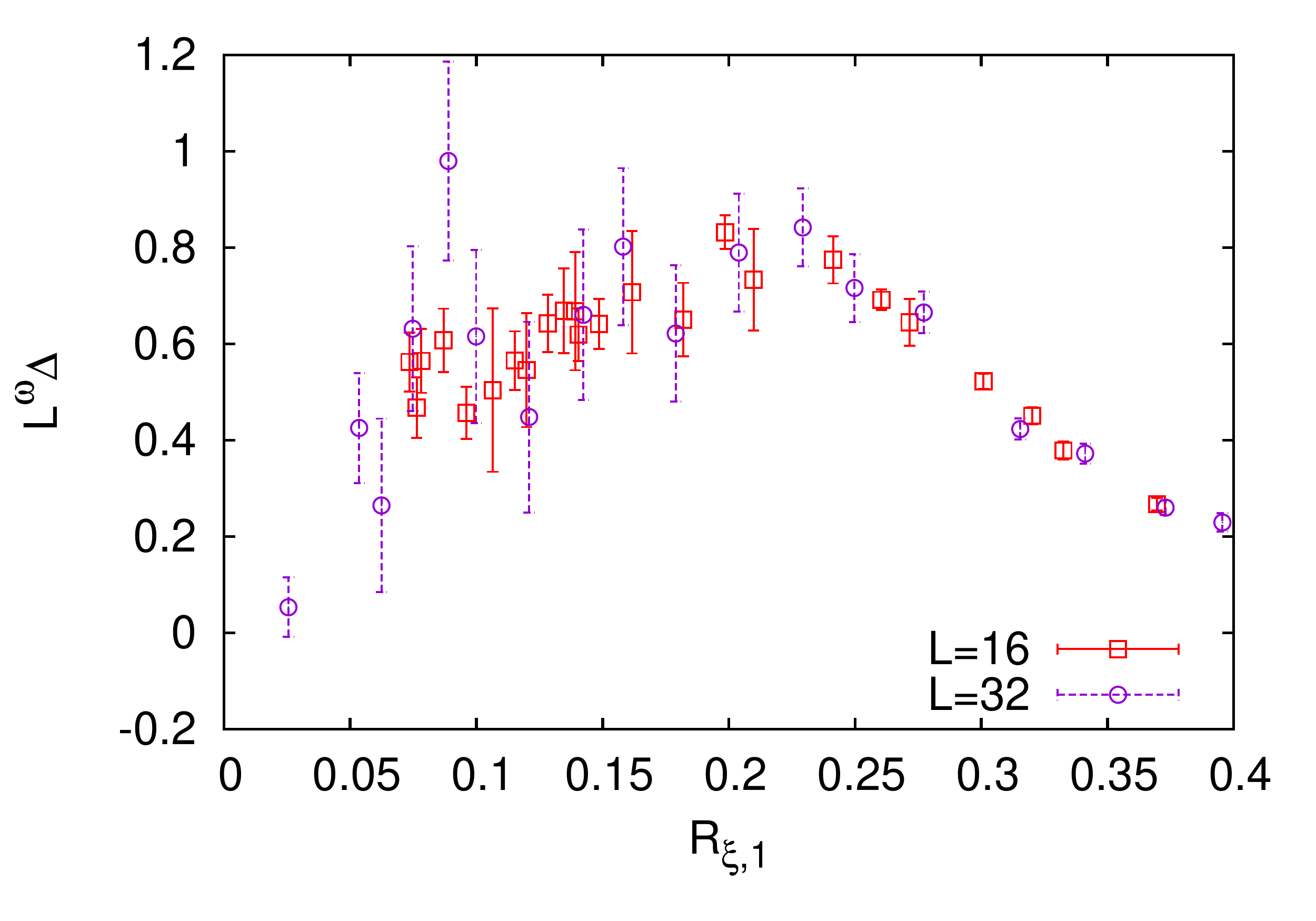}
\includegraphics*[scale=\graphicscale,angle=-90]{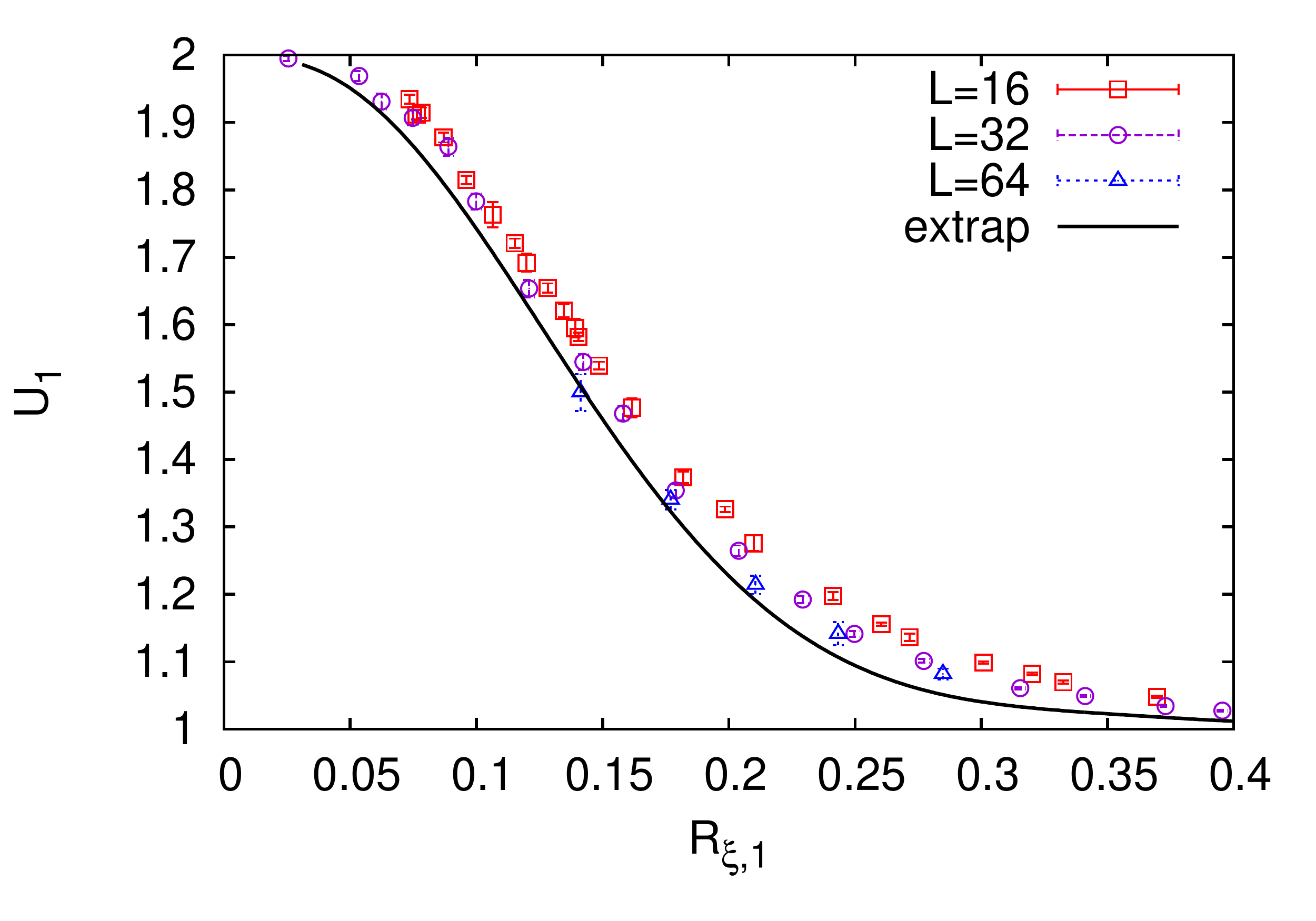}
\caption{ Bottom: Data of $U_1$ versus $R_{\xi,1}$ for the ${\mathbb
    Z}_{20}$ model and extrapolated scaling function
  $F_{\xi,1}(R_{\xi,1})$.  Top: Plot of $L^{\omega}\Delta$ versus
  $R_{\xi,1}$ where $\Delta$ is defined in Eq.~(\ref{deviations-def})
  and $\omega = 0.789$ is the correction-to-scaling exponent for the
  XY universality class \cite{Hasenbusch-19}.  }
\label{URxi1-Z20}
\end{figure}

\begin{figure}[tbp]
\includegraphics*[scale=\graphicscale,angle=-90]{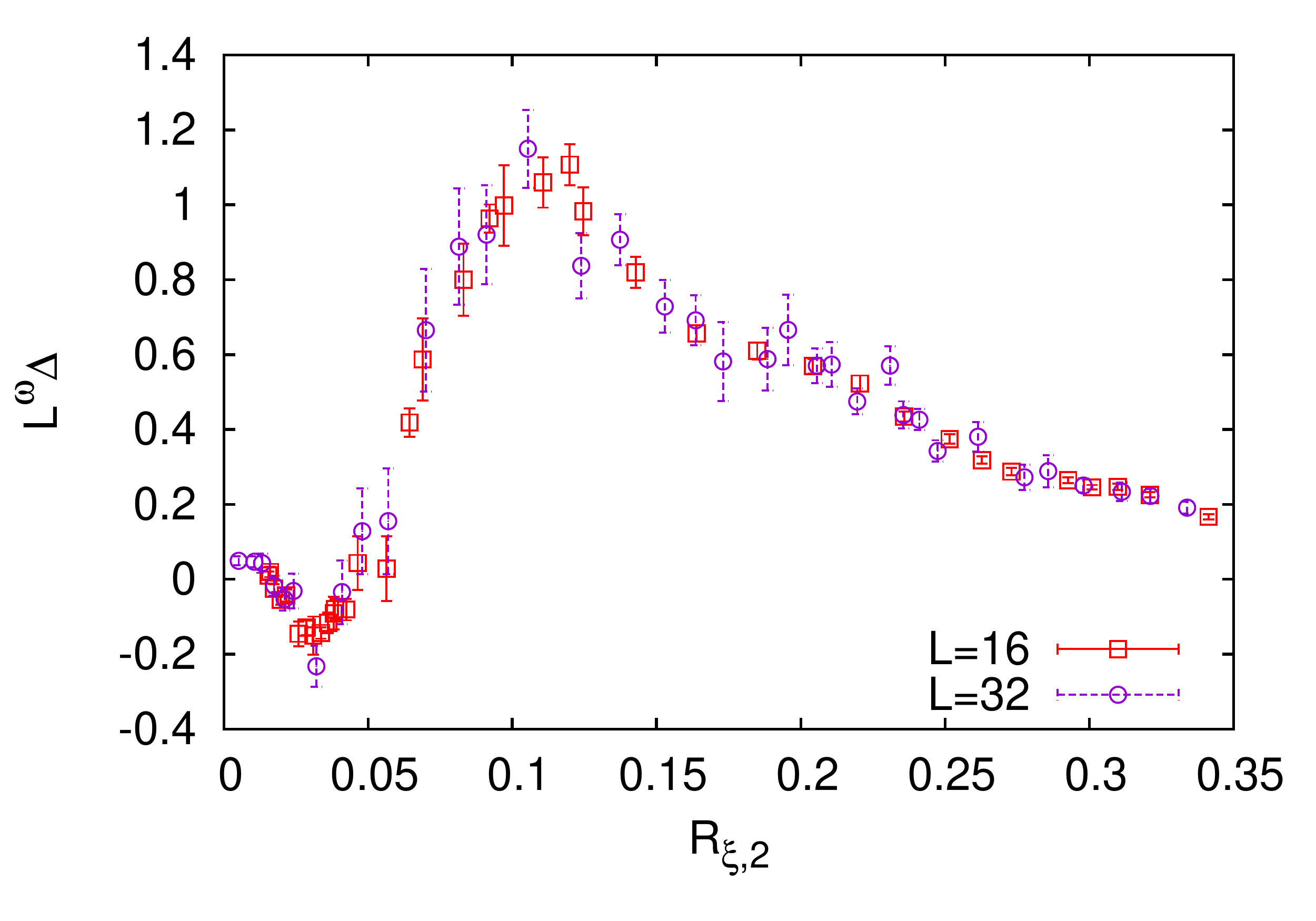}
\includegraphics*[scale=\graphicscale,angle=-90]{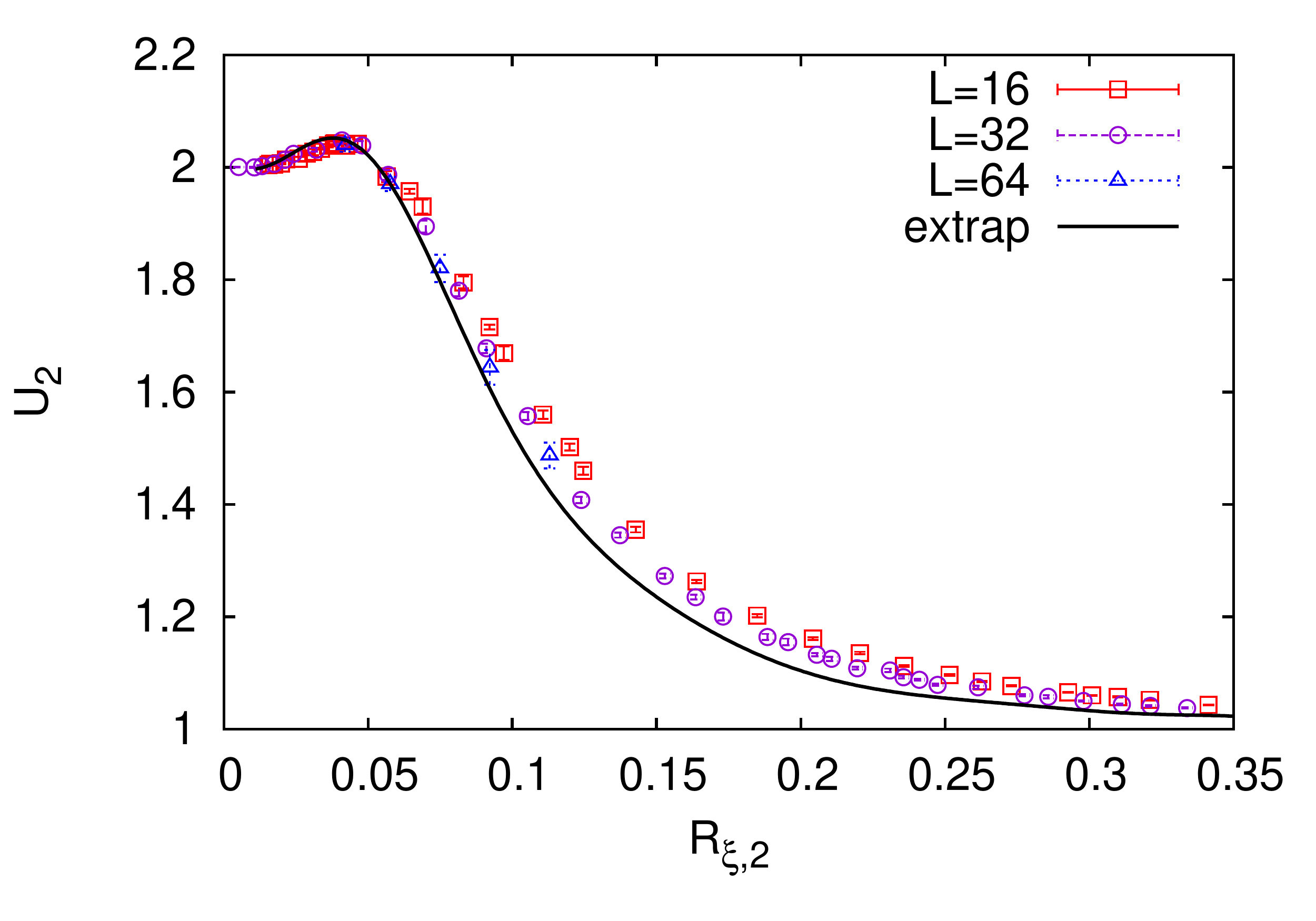}
\caption{ Bottom: Estimates of $U_2$ versus $R_{\xi,2}$ for the
  ${\mathbb Z}_{20}$ model and extrapolated scaling function
  $F_{\xi,2}(R_{\xi,2})$.  Top: $L^{\omega}\Delta$ versus
  $R_{\xi,2}$ where $\Delta$ is defined in Eq.~(\ref{deviations-def})
  and $\omega = 0.789$ is the correction-to-scaling exponent for the
  XY universality class \cite{Hasenbusch-19}.  }
\label{URxi2-Z20}
\end{figure}

Let us now discuss the computation of the scaling functions that
express $U_Q$ as a function of $R_{\xi,Q}$ for the XY universality
class.  To speed up the calculation, we have considered the ${\mathbb
  Z}_{20}$ clock model and we have performed extensive simulations on
lattices of size up to $L=64$. We report here the calculation of the
scaling functions for $Q=1$ and $Q=2$.  Also in this case, corrections
to scaling are sizeable, and therefore we have applied the same
strategy used in the Ising case.  The scaling functions have been
parametrized using polynomials and we have used $\omega = 0.789$
\cite{Hasenbusch-19}.  Results are reported in Figs.~\ref{URxi1-Z20}
and \ref{URxi2-Z20}, together with a scaling plot of the deviations.
Deviations scale nicely, confirming the reliability of the asymptotic
curves.  We have also determined the values of the two parameters at
the transition:
\begin{eqnarray}
   && U_1^* = 1.84(2)\,, \qquad R_{\xi,1}^* = 0.087(2)\,, \\
   && U_2^* = 2.02(1)\,, \qquad R_{\xi,2}^* = 0.022(2)\,. 
\end{eqnarray}

The results we have obtained for the Ising and XY model are the
relevant ones that shall be compared with the numerical data for the
gauge-scalar model.  There are, however, a few subtleties that should
be taken into account.

First, in the ${\mathbb Z}_4$ clock model, although the transition
belongs to the Ising universality class, the relation between
${\mathbb Z}_4$ and Ising correlation functions and Binder parameters
is not trivial, as discussed in App.~\ref{App.A}. Relation
(\ref{binderZ4_Q1}) allows us to relate the Binder parameter $U_1$ in
the ${\mathbb Z}_4$ model in terms of the Ising Binder parameter. We
have also determined the ${\mathbb Z}_4$ Binder parameter $U_2$ and
$R_{\xi,2}$ that are associated with Ising replica correlations. At
the critical point, performing the same analysis we did in the Ising
case, we obtain
\begin{equation}
U_2^* = 3.09(1)\,, \qquad R_{\xi,2}^* = 0.032(1)\,.
\end{equation}
For $q \ge 5$, the transition in the ${\mathbb Z}_q$ clock model
belongs to the XY universality class. However, this does not imply
that all correlation functions $G_Q({\bm x},{\bm y})$ are the same in
the ${\mathbb Z}_q$ clock model and in the XY model.  For instance, in
the ${\mathbb Z}_q$ model we have the relations
\begin{equation}
     G_Q({\bm x},{\bm y}) = G_{Q'}({\bm x},{\bm y})\,, 
\qquad Q' = |Q - n q|\,,
\end{equation}
for any integer $n$. These relations do not hold in the XY model.
Similar relations hold for the Binder parameters $U_Q$. We have
studied this issue in the ${\mathbb Z}_6$ clock model, verifying in
this case that the scaling function of $U_Q$ versus $R_{\xi,Q}$ is the
same in the ${\mathbb Z}_6$ and in the XY model for $Q=1,2$, while it
differs for $Q = 3$. The different behavior can be easily proved by
noting that, in the disordered limit ($R_{\xi,3}\to0$), we have $U_3 =
3$, 2 in the ${\mathbb Z}_6$ model and in the XY model,
respectively. Similar arguments can be used for any $q$, to show that
$U_Q$ differs in the ${\mathbb Z}_q$ model and in the XY model for $Q
\ge q/2$.


\begin{thebibliography}{99}

\bibitem{Fradkin_book} 
E. Fradkin, {\em Field Theories of Condensed Matter Physics},
2nd ed. (Cambridge Univ. Press, Cambridge, 2013). 

\bibitem{MM_book}
R. Moessner and J. E. Moore, 
{\em Topological Phases of Matter},
(Cambridge Univ. Press, Cambridge, 2021). 

\bibitem{RS-90} N. Read and S. Sachdev, Spin-Peierls, valence-bond
  solid, and N\'eel ground states of low-dimensional quantum
  antiferromagnets, Phys. Rev. B {\bf 42}, 4568 (1990).

\bibitem{SBSVF-04}
T. Senthil, L. Balents, S. Sachdev, A. Vishwanath, and M. P. A. Fisher,
Quantum Criticality beyond the Landau-Ginzburg-Wilson Paradigm,
Phys. Rev. B {\bf 70}, 144407 (2004).

\bibitem{TIM-05} S. Takashima, I. Ichinose, and T. Matsui, CP$^1$+U(1)
  lattice gauge theory in three dimensions: Phase structure, spins,
  gauge bosons, and instantons, Phys. Rev. B {\bf 72}, 075112 (2005).

\bibitem{TIM-06} S. Takashima, I. Ichinose, and T. Matsui,
  Deconfinement of spinons on critical points: Multiflavor CP$^1$+U(1)
  lattice gauge theory in three dimension, Phys. Rev. B {\bf 73},
  075119 (2006).

\bibitem{Kaul-12} R. K. Kaul, Quantum phase transitions in bilayer
  SU($N$) antiferromagnets, Phys. Rev. B {\bf 85}, 180411(R) (2012).

\bibitem{KS-12} R. K. Kaul and A. W. Sandvik, Lattice Model for the
  SU($N$) N\'eel to Valence-Bond Solid Quantum Phase Transition at
  Large $N$, Phys. Rev. Lett. {\bf 108}, 137201 (2012).

\bibitem{BS-13} 
T. A. Bojesen and A. Sudb\o, 
Berry phases, current lattices, and suppression of phase transitions in 
a lattice gauge theory of quantum antiferromagnets, 
Phys. Rev. B {\bf 88}, 094412 (2013).

\bibitem{BMK-13} M. S. Block, R. G. Melko, and R. K. Kaul, Fate of
  CP$^{N-1}$ fixed point with $q$ monopoles, Phys. Rev. Lett. {\bf
    111}, 137202 (2013).

\bibitem{NCSOS-15} A. Nahum, J. T. Chalker, P. Serna, M. Ortu\`no, and
  A. M. Somoza, Deconfined Quantum Criticality, Scaling Violations,
  and Classical Loop Models, Phys. Rev. X {\bf 5}, 041048 (2015).

\bibitem{WNMXS-17} C. Wang, A. Nahum, M. A. Metliski, C. Xu, and
  T. Senthil, Deconfined Quantum Critical Points: Symmetries and
  Dualities, Phys. Rev. X {\bf 7}, 031051 (2017).

\bibitem{Sachdev-19}
S. Sachdev,
Topological order, emergent gauge fields, and Fermi surface reconstruction,
Rep. Prog. Phys. {\bf 82}, 014001 (2019).


\bibitem{FS-79}
E. Fradkin and S. H. Shenker, 
Phase diagrams of lattice gauge theories with Higgs fields,
Phys. Rev. D {\bf 19}, 3682 (1979).

\bibitem{BPPW-81}
K. C. Bowler, G. S. Pawley, B. J. Pendleton, and D. J. Wallace, 
Phase diagrams of U(1) lattice Higgs models,
Phys. Lett. B {\bf 104}, 481 (1981).

\bibitem{PV-19-AH3d}
A. Pelissetto and E. Vicari,
Multicomponent compact Abelian-Higgs lattice models, 
Phys. Rev. E {\bf 100}, 042134 (2019).

\bibitem{BPV-20-AHq2}
C. Bonati, A. Pelissetto, and E. Vicari,
Higher-charge three-dimensional compact lattice Abelian-Higgs models, 
Phys. Rev. E {\bf 102}, 062151 (2020).

\bibitem{BPV-22-AHq}
C.~Bonati, A.~Pelissetto, and E.~Vicari,
Critical behaviors of lattice U(1) gauge models and three-dimensional
Abelian-Higgs gauge field theory,
Phys. Rev. B {\bf 105}, 085112 (2022).

\bibitem{PV-20-mfcp} 
A. Pelissetto and E. Vicari, 
Three-dimensional monopole-free CP$^{N-1}$ models, 
Phys. Rev. E {\bf 101}, 062136 (2020).

\bibitem{BPV-22-mfcp} 
C. Bonati, A. Pelissetto, and E. Vicari, 
Three-dimensional monopole-free CP$^{N-1}$ models:
Behavior in the presence of a quartic potential,
arXiv:2202.04614.

\bibitem{NCSOS-11}
A. Nahum, J. T. Chalker, P. Serna, M. Ortu\~no, and A. M. Somoza,
3D Loop Models and the CP$^{N-1}$ $\sigma$ Model,
Phys.  Rev.  Lett. {\bf 107}, 110601 (2011).

\bibitem{NCSOS-13}
A. Nahum, J. T. Chalker, P. Serna, M. Ortu\~no, and A. M. Somoza,
Phase transitions in three-dimensional loop models and the CP$^{N-1}$ 
$\sigma$ model, Phys. Rev. B {\bf 88}, 134411 (2013).

\bibitem{BPV-21-NCQED} 
C.~Bonati, A.~Pelissetto, and E.~Vicari, 
Lattice Abelian-Higgs model with noncompact gauge fields, 
Phys. Rev. B {\bf 103}, 085104 (2021).

\bibitem{HLM-74} 
B. I. Halperin, T. C. Lubensky, and S. K. Ma,
  First-Order Phase Transitions in Superconductors and Smectic-A
  Liquid Crystals, Phys. Rev. Lett. {\bf 32}, 292 (1974).

\bibitem{FH-96} 
R. Folk and Y. Holovatch, On the critical fluctuations
  in superconductors, J. Phys. A {\bf 29}, 3409 (1996).

\bibitem{IZMHS-19} 
B. Ihrig, N. Zerf, P. Marquard, I. F. Herbut, and
  M. M. Scherer, Abelian Higgs model at four loops, fixed-point
  collision and deconfined criticality, 
Phys. Rev. B {\bf 100}, 134507 (2019).

\bibitem{Wegner-71}
F. J. Wegner, 
Duality in generalized Ising models and phase transitions without
local order parameters, 
J. Math. Phys. {\bf 12}, 2299 (1971).

\bibitem{Kogut-79}
J. B. Kogut,
An introduction to lattice gauge theory and spin systems,
Rev. Mod. Phys. {\bf 51}, 659 (1979).

\bibitem{GR-02}
F. Gliozzi and A. Rago,
Monopole clusters, center vortices, and confinement in a 
$Z_2$ gauge-Higgs system,
Phys. Rev. D {\bf 66}, 074511 (2002).

\bibitem{GGRT-03} L. Genovese, F. Gliozzi, A. Rago, and C. Torrero,
  The phase diagram of the three-dimensional $Z_2$ gauge Higgs system
  at zero and finite temperature, Nucl. Phys. B (Proc. Suppl.) 
  {\bf 119}, 894 (2003).

\bibitem{SF-00}
T. Senthil and  M. P. A. Fisher, 
${\mathbb Z}_2$ gauge theory of electron fractionalization in
strongly correlated systems, 
Phys. Rev. B {\bf 62}, 7850 (2000).

\bibitem{MSF-01}
R. Moessner, S. L. Sondhi, and E. Fradkin,
Short-ranged resonating valence bond physics, quantum dimer models, 
and Ising gauge theories,
Phys. Rev. B {\bf 65}, 024504 (2001).

\bibitem{SSS-02}
R. D. Sedgewick, D. J. Scalapino, and R. L. Sugar, 
Fractionalized phase in an XY-$Z_2$ gauge model, 
Phys. Rev. B {\bf 65}, 054508 (2002).

\bibitem{PD-05}
D. Podolsky and E. Demler, 
Properties and detection of spin nematic order in
strongly correlated electron systems, 
New J. Phys. {\bf 7}, 59 (2005).

\bibitem{Nussinov-05} Z. Nussinov, Derivation of the Fradkin-Shenker
  result from duality: Links to spin systems in external magnetic
  fields and percolation crossovers, Phys. Rev. D {\bf 72}, 054509
  (2005).

\bibitem{LRT-93}
P. E. Lammert, D. S. Rokhsar, and J. Toner,
Topology and nematic ordering, Phys. Rev. Lett. {\bf 70}, 1650 (1993);
Topology and nematic ordering. I. A gauge theory, 
Phys. Rev. E {\bf 52}, 1778 (1995); 
J. Toner, P. E. Lammert, and D. S. Rokhsar,
Topology and nematic ordering. II. Observable critical behavior,
Phys. Rev. E {\bf 52}, 1801 (1995).

\bibitem{LNNSWZ-15}
K. Liu, J. Nissinen, Z. Nussinov, R.-J. Slager, K. Wu, and J. Zaanen,
Classification of nematic order in 2+1 dimensions: Dislocation melting in an 
$O(2)/{\mathbb Z}_N$ lattice gauge theory,
Phys. Rev. B {\bf 91}, 075103 (2015).


\bibitem{VDS-09} J. Vidal, S. Dusuel, and K. P. Schmidt, Low-energy
  effective theory of the toric code model in a parallel magnetic
  field, Phys. Rev. B {\bf 79}, 033109 (2009).

\bibitem{TKPS-10} I. S. Tupitsyn, A. Kitaev, N. V. Prokofev, and
  P. C. E. Stamp, Topological multicritical point in the phase diagram
  of the toric code model and three-dimensional lattice gauge Higgs
  model, Phys. Rev. B {\bf 82}, 085114 (2010).

\bibitem{WDP-12} F. Wu, Y. Deng, and N. Prokofev, Phase diagram of the
  toric code model in a parallel magnetic field, Phys. Rev. B {\bf
    85}, 195104 (2012).

\bibitem{SSN-21} A. Somoza, P. Serna, and A. Nahum, Self-dual
  criticality in three-dimensional ${\mathbb Z}_2$ gauge theory with
  matter, Phys. Rev. X {\bf 11}, 041008 (2021).

\bibitem{Grady-21} M. Grady, Exploring the 3D Ising gauge-Higgs model
  in exact Coulomb gauge and with a gauge-invariant substitute for
  Landau gauge, arXiv:2109.04560.

\bibitem{HSAFG-21} L. Homeier, C. Schweizer, M. Aidelsburger,
  A. Fedorov, and F. Grusdt, ${\mathbb Z}_2$ lattice gauge theories
  and Kitaev's toric code: A scheme for analog quantum simulation,
Phys. Rev. B {\bf 104}, 085138 (2021). 

\bibitem{BPV-21-Z2Higgs}
C. Bonati, A. Pelissetto, and E. Vicari,
Multicritical point of the three-dimensional
${\mathbb Z}_2$ gauge Higgs model,
arXiv:2112.01824

\bibitem{SSNHS-03} J. Smiseth, E. Sm{\o}rgrav, F. S. Nogueira,
  J. Hove, and A. Sudb{\o}, Phase Structure of $d = 2 + 1$ Compact
  Lattice Gauge Theories and the Transition from Mott Insulator to
  Fractionalized Insulator, Phys. Rev. B {\bf 67}, 205104 (2003).

\bibitem{HS-03} J. Hove and A. Sudb{\o}, Criticality versus $q$ in the
  (2+1)-dimensional $Z_q$ clock model, Phys. Rev. E {\bf 68}, 046107
  (2003)

\bibitem{SGS-20}
H. Shao, W. Guo, A. W. Sandvik, Monte Carlo renormalization flows in the
space of relevant and irrelevant operators: Application to three-dimensional
clock models, Phys. Rev. Lett. 124 080602 (2020).

\bibitem{PSS-20} P. Patil, H. Shao, and A. W. Sandvik, Unconventional
  U(1) to $Z_q$ cross-over in quantum and classical q-state clock
  models, Phys. Rev. B {\bf 103}, 054418 (2021).

\bibitem{BSADGG-19}
L. Barbiero, C. Schweizer, M. Aidelsburger, E. Demler, N. Goldman, and
F. Grusdt,
Coupling ultracold matter to dynamical gauge fields in optical lattices: From
flux attachment to $Z_2$ lattice gauge theories,
Science Adv. {\bf 5}, eaav7444 (2019).

\bibitem{Hasenbusch-19} M. Hasenbusch, Monte Carlo study of an
  improved clock model in three dimensions, Phys. Rev. B {\bf 100},
  224517 (2019).

\bibitem{Hasenbusch-20}
M. Hasenbusch, 
Monte Carlo study of a generalized icosahedral model on the simple cubic
lattice,
Phys. Rev. B {\bf 102}, 024406 (2020).

\bibitem{KS-19}
N. Klco and M. J. Savage,
Digitization of scalar fields for quantum computing,
Phys. Rev. A {\bf 99}, 052335 (2019).

\bibitem{JLZ-20}
Y. Ji, H. Lamm, and S. Zhu,
Gluon field digitization via group space decimation for quantum
computers,
Phys. Rev. D {\bf 102}, 114513 (2020).

\bibitem{CLB-86} M. S. S. Challa, D. P. Landau, and K. Binder,
  Finite-size effects at temperature-driven first-order transitions,
  Phys. Rev. B {\bf 34}, 1841 (1986).

\bibitem{VRSB-93} K.~Vollmayr, J. D.~Reger, M.~Scheucher, and
  K.~Binder, Finite size effects at thermally-driven first order phase
  transitions: A phenomenological theory of the order parameter
  distribution, Z.~Phys.~B {\bf 91}, 113 (1993).

\bibitem{CPV-03}
P. Calabrese, A. Pelissetto, and E. Vicari,
Multicritical Phenomena in $O(n_1)\oplus O(n_2)$ Symmetric Theories,
Phys. Rev. B {\bf 67}, 054505 (2003).

\bibitem{PAD-15}
S. Pujari, F. Alet, and K. Damle, Transitions to valence-bond solid order
in a honeycomb lattice antiferromagnet, Phys. Rev. B 91, 104411 (2015)

\bibitem{KPSV-16}
F. Kos, D. Poland, D. Simmons-Duffin, and A. Vichi,
J. High. Energy Phys. {\bf 08} (2016) 036.

\bibitem{BCCGPS-14} O.~Borisenko, V.~Chelnokov, G.~Cortese,
  M.~Gravina, A.~Papa, and I.~Surzhikov, Critical behavior of 3D Z(N)
  lattice gauge theories at zero temperature, Nucl. Phys. B
  \textbf{879}, 80 (2014).

\bibitem{FXL-18} A. M. Ferrenberg, J. Xu, and D. P. Landau, Pushing
  the limits of Monte Carlo simulations for the three-dimensional
  Ising model, Phys. Rev. E {\bf 97}, 043301 (2018).

\bibitem{NRR-03}
T. Neuhaus, A. Rajantie, and K. Rummukainen,
Numerical study of duality and universality in a frozen superconductor,
Phys. Rev. B {\bf 67}, 014525 (2003).

\bibitem{EPPSSV-12-14}
S. El-Showk, M. F. Paulos, D. Poland, S. Rychkov, D. Simmons-Duffin, 
and A.  Vichi, Phys. Rev. D {\bf 86}, 025022 (2012);
J. Stat. Phys. {\bf 157},869 (2014)

\bibitem{CLLPSSV-20} S. M. Chester, W. Landry, J. Liu, D. Poland,
  D. Simmons-Duffin, N. Su, and A. Vichi, Carving out OPE space and
  precise O(2) model critical exponents, J. High Energy Phys. {\bf
    06}, 142 (2020).

\bibitem{CPV-00}
J. M. Carmona, A. Pelissetto, and E. Vicari,
$N$-component Ginzburg-Landau Hamiltonian 
with cubic anisotropy: A six-loop study, 
Phys. Rev. B {\bf 61}, 15136 (2000).
\end{thebibliography}
\end{document}